\documentclass{aa}
\usepackage{graphicx}
\usepackage{epstopdf}
\usepackage{amssymb,amsmath}
\usepackage[colorlinks]{hyperref}
\usepackage{aalongtable}
\usepackage{lscape}
\usepackage{color}
\usepackage{natbib}
\usepackage{enumitem}
\usepackage{ulem}
\usepackage{listings}
\newcommand{\foothref}[1]{\footnote{\url{#1}}} % for URLs in footnotes: make them clickable (href)
\newcommand{\gr}{{$\gamma$-ray}}
\newcommand{\lsim}{{\lower.5ex\hbox{$\; \buildrel < \over \sim \;$}}}
\newcommand{\gsim}{{\lower.5ex\hbox{$\; \buildrel > \over \sim \;$}}}

\newcommand{\nufnu}{$\nu$f$({\nu})$}

\newcommand{\ergs}{erg cm$^{-2}$ s$^{-1}$}

\newcommand{\swiftdeepsky}{\textit{Swift-DeepSky}} % This is the *name* of the software
\begin{document}
   \title{Open Universe for Blazars: a new generation of astronomical products based on 14 years of Swift-XRT data.}
   \author{
          P. Giommi   \inst{1,2,3}
          \and  C.H. Brandt  \inst{3,4}
          \and  U. Barres de Almeida \inst{5,3}
          \and  A.M.T. Pollock \inst{6}
          \and  F. Arneodo \inst{7}
          \and  Y. L. Chang  \inst{3}
          \and  O. Civitarese\inst{8,9}
          \and  M. De Angelis\inst{1}
          \and 	V. D'Elia \inst{10,12} 
          \and J. Del Rio Vera\inst{11}
          \and S. Di Pippo\inst{11}
          \and R. Middei \inst{13}
          \and A. V. Penacchioni\inst{8}
	      \and M. Perri   \inst{10,12} 
	      \and R. Ruffini \inst{3}
	      \and N. Sahakyan\inst{14}
	      \and S. Turriziani\inst{15}
         }        
        \institute{Italian Space Agency, ASI, via del Politecnico snc, 00133 Roma, Italy
         \and Institute for Advanced Study, Technische Universit{\"a}t M{\"u}nchen, Lichtenbergstrasse 2a, D-85748 Garching bei M\"unchen, Germany
         \and ICRANet, P.zza della Repubblica 10, 65122, Pescara, Italy
         \and Jacobs University, Physics and Earth Sciences, Campus Ring 1, 28759, Bremen, Germany  
         \and Centro Brasileiro de Pesquisas F\'isicas, Rua Dr. Xavier Sigaud 150, 22290-180, Rio de Janeiro, Brazil
         \and Department of Physics and Astronomy, University of Sheffield, Hounsfield Road, Sheffield S3 7RH, England
         \and New York University Abu Dhabi, Abu Dhabi, UAE
%        \and Sapienza Universit\'a di Roma, IcraNet, Dipartimento di Fisica, Piazzale Aldo Moro 5, I-00185 Roma, Italy
         \and Institute of Physics. IFLP-CONICET. diag 113 e/63-64.(1900) La Plata. Argentina
         \and Department of Physics. University of La Plata. 49 and 115.C.C.67 (1900) La Plata, Argentina
         \and Space Science Data Center, SSDC, ASI, via del Politecnico snc, 00133 Roma, Italy
         \and United Nations Office for Outer Space Affairs, UNOOSA, Vienna, Austria
         \and INAF - Osservatorio Astronomico di Roma, via Frascati 33, I-00040 Monteporzio Catone, Italy
        %  \and CAPES Foundation, Ministry of Education of Brazil, Bras\'ilia, Brazil
         \and Dipartimento di Matematica e Fisica, Universit\`a degli studi Roma Tre, Via della Vasca Navale 84, 00146, Roma, Italy 
         \and ICRANet-Armenia, Marshall Baghramian Avenue 24a, Yerevan 0019, Armenia.
         \and Computational Astrophysics Laboratory - RIKEN, 2-1 Hirosawa, Wako, Saitama 351-0198, Japan
         \\
         \email{giommipaolo@gmail.com}
        }
\abstract
{}% leave it empty if necessary  
{
Open Universe for blazars is a set of high-transparency multi-frequency data products for blazar science, and the tools designed to generate them. Blazars are drawing growing interest following the consolidation of their position as the most abundant type of source in the extragalactic very-high energy \gr\ sky, and because of their status as prime candidate sources in the nascent field of multi-messenger astrophysics. As such, blazar astrophysics is becoming increasingly data driven, depending on the integration and combined analysis of large quantities of data from the entire span of observational astrophysics techniques. The project was therefore chosen as one of the pilot activities within the United Nations Open Universe Initiative.
} 
{
We aim to deliver innovative data science tools for multi-messenger astrophysics. In this work we developed a data analysis pipeline called \swiftdeepsky\, based on the Swift XRTDAS software and the XIMAGE package, encapsulated into a Docker container. \swiftdeepsky\, downloads and reads low-level data, generates higher-level products, detects X-ray sources and estimates several intensity and spectral parameters for each detection, thus facilitating the generation of complete and up-to-date science-ready catalogues from an entire space-mission dataset. The Docker version of the pipeline -- whose concept can be reproduced with other missions -- and its derived products is publicly available from the Open Universe Website at openuniverse.asi.it 
%}
} 
{
%As a first application of our innovative approach, 
We present the results of a detailed X-ray image analysis based on \swiftdeepsky\, that was run on all Swift XRT observations including a known blazar, carried out during the first 14 years of operations of the Neil Gehrels Swift Observatory.
Short exposures executed within one week of each other have been added to increase sensitivity, which ranges between $\sim 1\times 10^{-12}$ and $\sim1\times 10^{-14}$ \ergs (0.3-10.0~keV). 
After cleaning for problematic fields, the resulting database includes over 27,000 images integrated in different X-ray bands, and a catalogue, called 1OUSXB, that provides intensity and spectral information for 33,396 X-ray sources, 8,896 of which are single or multiple detections of 2,308 distinct blazars. All the results can be accessed on-line in a variety of ways: e.g., from the Open Universe portal at openuniverse.asi.it, through Virtual Observatory services, via the VOU-Blazar tool and the SSDC SED builder. 
One of the most innovative aspects of this work is that the results can be safely reproduced and extended by anyone.
}
{}
\titlerunning{a new generation of astronomical products based on 14 years of Swift-XRT data}
\authorrunning{Giommi, P. et al.}
\maketitle
%
%________________________________________________________________
\section{Introduction}\label{intro}

Providing open and transparent access to scientific data is a social obligation of the scientific community, and central to the progress of research in the era of big data and of data-driven science. Facilitated access to scientific data can also play a critical role in education, capacity building, and in the promotion of citizen science. For this reason the United Nations recognised that free access to space-science data is a strategic objective contributing to the achievement of the Sustainable Development Goals (SDGs)\foothref{http://www.unoosa.org/oosa/oosadoc/data/documents/2018/aac.105/aac.1051175_0.html}.
%[cite A/AC.105/1175 or another more appropriate ref.]
%\sout{and a cultural strategic objective that contributes to the achievement of the United Nations Sustainable Development Goals.}

%[By Jorge]
%\textcolor{magenta} {
In the field of astronomy, the third United Nations Conference on the Exploration and the Peaceful Uses of Outer Space, UNISPACE III, requested action to be taken to improve scientific knowledge of near and outer space by promoting cooperative activities in such areas as astronomy, space biology and medicine, space physics, the study of near-Earth objects, and planetary exploration. This request is echoed in "Open Universe".
%}

"Open Universe"\foothref{http://openuniverse.asi.it} \citep{GiommiOU} is an initiative proposed by Italy to the United Nations Committee on the Peaceful Uses of Outer Space (COPUOS) in 2016\foothref{http://www.unoosa.org/res/oosadoc/data/documents/2016/aac_1052016crp/aac_1052016crp_6_0_html/AC105_2016_CRP06E.pdf}
%\textcolor{blue}{[cite A/AC./105/2016/CRP.6} 
that is now being actively developed by a number of countries in coordination with the UN Office for Outer Space Affairs (UNOOSA).
%{\color{magenta} By Sara T. I noticed that through the paper sometimes we write UNOOSA sometime UN-OOSA, maybe we should uniform to a unique spelling}
Its main goal is to contribute in making astronomy and space-science data
%\textcolor{red}{[U.B. ensure is a bit vague; maybe could rephrase using more operative verbs such as "develop, foster etc,..."]}
%P.G. I tried, but I still think that the word "ensure" is stronger than others...
%U.B. What about "contribute"? "Ensure" sounds a bit like we are saying Open Universe will guarantee or will be able to secure by itself this goal of open space science data. Contribute is less strong, and maybe fairer, as other works will also be contributing to the same goal...
%\sout{that astronomy and space science data become} 
more openly available, easily discoverable, free of bureaucratic, administrative or technical barriers, and therefore usable by the widest possible community, from professional researchers to all people interested in space science.
In doing so, Open Universe aims to support an increase in productivity of space research, facilitate the emerging field of data-driven science, and stimulate a significant acceleration towards the democratisation of space science and the benefits therefrom. 
%{\color{magenta} [By JorCe] ogmment by Sara T. the word science is repeated 3 times in the last sentence. I'd suggest to remove the first one i.e. say only space research, and substitute the last one with knowledge, i.e.https://www.overleaf.com/2244812846ycsmtpcztqqd democratisation of knowledge instead of science}
%\textcolor{red}{U.B. The last sentence could be reformulated. It may sound a pretentious to the specialist reader of a technical publication. We can tone it down here and bring it back again in the conclusions, with arguments coming from the concrete example shown, of OUblazars.}

Open access to scientific data, and its associated technologies, has been actively addressed by a number of initiatives working in complementary ways, such as
%\sout{ is a subject that is actively discussed and addressed in various contexts, especially in the scientific sector. Consequently, a number of initiatives that approach this matter from different angles have emerged in recent years, like e.g.} 
the International Virtual Observatory Alliance (IVOA) \foothref{http://www.ivoa.net}, the European Open Science Cloud (EOSC)\foothref{https://www.eosc-portal.eu}, the Research Data Alliance (RDA)\foothref{https://www.rd-alliance.org}, the Astronomy ESFRI and Research Infrastructure Cluster (ASTERICS)\foothref{https://www.asterics2020.eu/}, among others. 

%\textcolor{blue}{
Starting from different angles and approaches, these various initiatives propose and develop solutions to support the reuse of astrophysical data, following general concepts which are now commonly accepted and widely endorsed, as expressed, for example, in the FAIR Guiding Principles for Scientific Data Management and Stewardship \citep{wilkinson2016fair} and in the particular application of the FAIR principles in astronomy and space science \citep{Pollock:2018}. 

%{\color{blue} 
The Open Universe Initiative stresses that open data by itself does not equal data transparency, which, in our view, requires compliance to all of the following criteria:
%\sout{science data can be considered \textit{transparent} if they} 
a) satisfy the FAIR principles;
b) optimise the scientific and sociological value of the data;
c) provide access via the simplest possible interfaces;
and d) require no additional processing.
%U.B. condition (b) sounds a bit confusing and abstract - difficult to understand what it means in practice that is different from FAIR; in (c) "other purposes" is too vague to be used in an operational definition; maybe we can be more specific / concrete?

%{\color{magenta} [by Sara:] add ASTERICS }
%{\color{blue} [by Carlos:] the Research Data Alliance (RDA) (https://www.rd-alliance.org).}

%\textcolor{red}{U.B. When we take into consideration all the initiatives that were cited, we are missing here to say what is special and different in Open Universe from all others, before we move to the specifics of OUblazars. Since our public will be of experts, and very critic, we need to get this sharp and right.}

%\sout{
In line with this general vision and, in particular, with the democratisation of science and its potential as a tool for development, Open Universe wishes to contribute to offering accessible space science data that can be used by anyone, and that can be flexibly employed in different applications from research to education, thus responding to the global demand for transparency of all goods produced through public funding. 
%\sout{to which Open Universe wishes to contribute by offering space science data for use by anyone,} 
%[By Jorge]
%\textcolor{magenta}{
Currently, the operational part of Open Universe can be accessed through its data portal (\url{openuniverse.asi.it}), which acts as a window to web-based initiatives collecting data. The portal provides also a few additional online functions for smart, purpose-built data mining and data visualisation, such as the "VOU-Blazars" and "VOU-SED" tools. The Initiative, however, is not limited to such services, but aims to the continuous development of new data products and data-oriented solutions. It is open to collaboration with all countries and institutions sharing our views on data transparency and open science, and willing to participate in activities to increase the reach of astronomical datasets, and lower the technical complexity required to work with them.
%}
%FA: perhaps "in a structured manner that increases...?"

%{\color{blue} 
As a first concrete application of this view, we present "Open Universe for Blazars", a project devoted to the generation of transparent scientific products from multi-frequency astronomical data by the Neil Gehrels Swift Observatory \citep[][hereafter Swift]{swift} and other astronomy satellites. 

This first project, developed as a pilot activity within the Initiative, focuses on blazars, astrophysical sources that are receiving increased attention as they are expected to abound in upcoming very high-energy \gr\ sky surveys\foothref{https://www.cta-observatory.org/wp-content/uploads/2017/11/ScienceWithCTA_FINAL2_updated.pdf}
\foothref{http://english.ihep.cas.cn/lhaaso}.
%{\color{red}[ref. missing]} 
In addition, blazars will likely play a crucial role in the nascent field of multi-messenger astrophysics, following their recent potential association with astrophysical high-energy neutrinos and Ultra High Energy Cosmic Rays \citep[UHECRs, e.g.][]{neutrino,Resconi2017}
%{\color{red}[ref. missing]}.

%{\color{red}This paragraph is misplaced. Once we start talking about OUBlazars, should not go back to general things}. 

%{\color{blue}
%\sout{In addition to the perceived relevance of the scientific topic chosen,}
The multi-messenger and big data characteristics of blazar research motivated the project as a test-bed for the incorporation of novel computational technologies to increase data transparency and usability. To that end, we pioneered the combination of Virtual Observatory services with Linux containers to provide enhanced accessibility to top-quality, reliable scientific data products on demand, with minimum recourse to mission-specific data reduction techniques, or computational knowledge by the user.

%\sout{To pursue our goal we take advantage of recent technological developments, including the availability of Linux Containers and Virtual Observatory services.} 
Linux Containers, and in particular the implementation by Docker Inc.\foothref{https://www.docker.com}, provide a portable solution for shipping software. 
%{\color{blue}
As shown below, this technology is suitable for a safe and robust distribution of ready-to-use complex astronomical data analysis packages, with a simple user interface, removing the need for a high-degree of specialised knowledge that would otherwise be required. This in turn opens the way to a number of flexible applications, such as in distributed and web-based intensive data analysis, mobile-compatible operation, and improved usability by non-experts for purposes of science, education or citizen science.

%\sout{This is particularly useful for scientific software, often requiting specialised expertise to setup, allowing us to distribute {\it ready-to-use} complex software packages.}

\section{Blazars and the Swift Observatory}\label{blazarsandswift}

Blazars are a special and uncommon type of Active Galactic Nucleus\footnote{only one every $\sim$ 350 objects listed in current catalogues of AGN is a blazar. See Sect. \ref{blazarlist}} \citep[AGN,][]{AGNReview}, distinguished by the emission of highly variable radiation across the entire electromagnetic spectrum.
It is commonly accepted that a large part of this radiation is generated by charged particles within a jet of material that moves away at relativistic speeds from the central supermassive black hole and happens to be pointing in the direction of the Earth \citep[see e.g.][]{Urry1995,Romero2017}. 

Blazars come in two sub-classes, FSRQs and BL Lacs, depending on their optical spectra: FSRQs show broad emission lines just like normal QSOs, while BL Lacs instead display at most weak emission lines, sometimes exhibit absorption features, and in many cases are completely featureless.
The observational appearance of blazars depends on a complex mix of physical and geometrical conditions, and their classification is 
subject to selection effects \citep{Giommi2012}.

The Spectral Energy Distribution (SED, a plot of the  \nufnu\  flux as a function of frequency $\nu$) of blazars has been studied in ever growing detail since the early days of multi-frequency astronomy
(see e.g. \cite{firstsedpaper,giommiansarimicol,fossati1998,ghisellini1998}, showing that it covers the entire electromagnetic spectrum and
and is characterised by a "double humped" shape, see Fig 1 of \cite{AGNReview}, or Fig. \ref{fig:SED-BLLac} and Fig. \ref{fig:SED-OJ287} below).
The low-energy component, peaking between the IR and the X-ray band, is generally attributed to synchrotron radiation produced by relativistic electrons moving in a magnetic field. The second component extends well into the \gr\ band, and is typically explained as either inverse Compton scattering of the electrons against photons, or other radiation mechanisms involving hadronic scenarios and neutrino emission.

Despite their rarity compared to other types of AGN, the particular radiation processes that cause blazars to emit across the entire electromagnetic spectrum and the particular geometrical circumstances that cause strong flux amplification, make these objects the most common type of extragalactic source so far detected in the microwave and \gr\ skies \citep{GiommiColafrancesco2004,3FGL,3LAC}. 
Blazars have often been associated with high-energy astrophysical neutrinos \citep[e.g.][]{Mannheim1995,AtoyanDermer2003,Tavecchio2014,Rodrigues2018,Dissecting}
and possibly to Ultra High Energy Cosmic Rays \citep[e.g.][]{Resconi2017}.
For the same reason, blazars are expected  to be found in large numbers in the high-energy extragalactic sky that will soon be surveyed by the new generation of powerful VHE observatories such as CTA, HAWK or LHAASO, and will likely play a crucial role in the emerging field of multi-messenger astrophysics.
%FA: Other? reference missing?
%U.B. I have some doubts about quoting LHAASO here - you wold have to mention HAWC, and its planned southern version, probably. But CTA's sensitivity is a factor of several 10x greater than LHAASO or these others...

%Making all the existing multi-frequency data available in a ready to use (transparent) way is necessary to match the needs of present and future astronomical research.

Blazar research has always been central to the scientific program of the Swift observatory \citep{swift}, which is a NASA mission launched in 2004 November and still fully operational. The Swift hardware and software were built by an international collaboration involving US, Italy, and United Kingdom. The original main scientific driver of the Swift mission was to detect gamma-ray bursts (GRBs) in the hard X-ray band with the Burst Alert Telescope \citep[BAT,][]{bat}, and quickly follow-up their emission at longer wavelength with the X-Ray Telescope \citep[XRT,][]{Burrows2005}, and the Ultraviolet/Optical Telescope \citep[UVOT,][]{uvot}.
Although specifically designed to carry out GRB science, Swift has proven to be an extremely effective multi-purpose multi-frequency observatory.
During its first 14 years of operations, the satellite carried out over 24,000 pointings \footnote{here we consider only single  observations\_ids with XRT exposures larger than 200 seconds in PC readout mode} 
%\textcolor{blue}{Comment by Matteo: it should be specified if the number of XRT exsposures refers only to PC mode exposures or PC+WT}}
of approximately 2,600 catalogued blazars, for a total of 44.6 Ms of XRT net exposure time, corresponding to  $\approx$ 14\% of the entire mission scientific program. The XRT is an X-ray telescope with a CCD at its focus that is normally operated in either of two readout modes, Photon Counting (PC) mode, which provides full imaging and spectral information, routinely used to observe targets that are faint or moderately bright X-ray sources, and Windowed Timing (WT) mode, that does not provide full imaging information and is used to measure the X-ray spectrum and the time variability of sources brighter than $\approx 2\times10^{-11}$ erg cm$^{-2}$ s$^{-1}$ in the 0.3-10.0 keV band.
% {\color{red}(38.2 Ms in PC mode and 5.3 Ms in WT mode )}.

\begin{figure}[h]
\centering
\includegraphics[width=9.0cm]{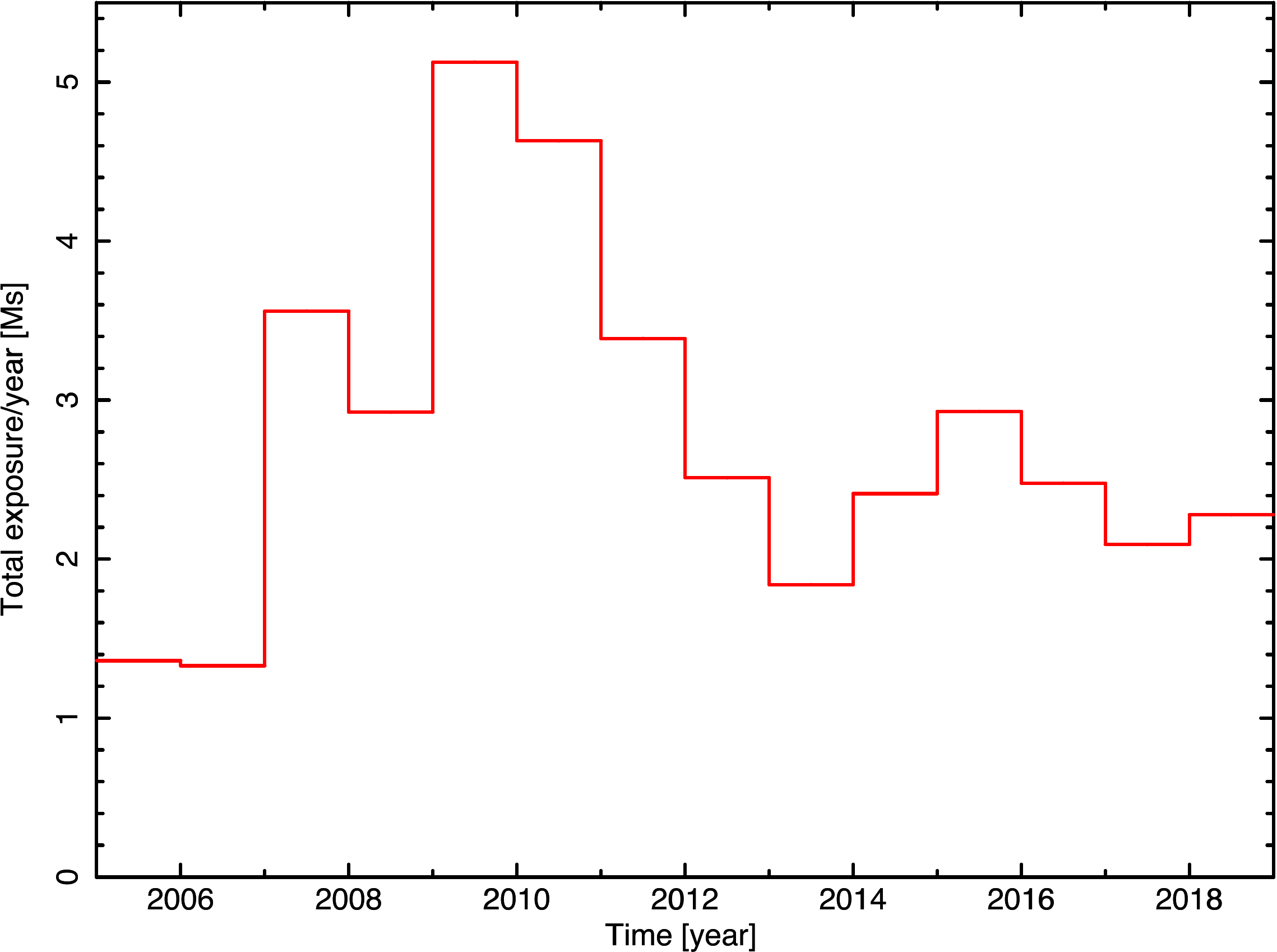}
\caption{The distribution of XRT exposure time in PC mode dedicated to known blazars over the 14 years of Swift mission operations covered by this work.}
\label{fig:year_expo}       
\end{figure}

\begin{figure}[h]
\centering
\includegraphics[width=9.0cm]{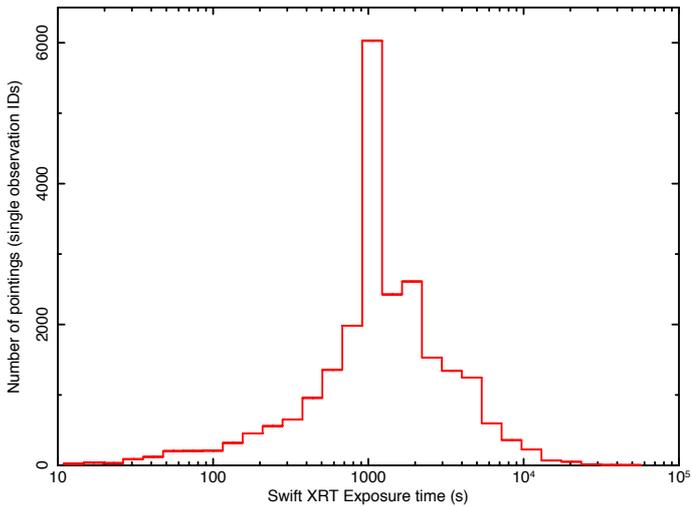}
\caption{The distribution of Swift exposure times of single Observation\_ID pointings  where the XRT field of view includes a known blazar. 
Note the strong peak at 1~ks.
%{\color{red}This figure seems quite different from the analogous figure 2 of Evans et al. 2014. Matteo, can you help me understand why they do not have the strong peak at expo=1000?}
}
\label{fig:expo_distr}       
\end{figure}

Fig. \ref{fig:year_expo} plots the amount of XRT exposure time dedicated to observations of blazars in PC mode during the first 14 years of Swift operations. %ranging between about 1 and 3 Ms/year, with a peak of $\sim $4.3 Ms in the period 2009-2010. 
The distribution of exposure times of individual XRT pointings, each characterised by an 11 digit unique identifier called Observation\_ID or sequence number, ranges from a few hundred seconds to about 10~ks, with a clear peak at 1~ks  (see fig. \ref{fig:expo_distr}).  

All Swift data immediately enter the public domain and are openly available to anyone. The official Swift archives provide access to low and intermediate level data products together with complex reduction and analysis software and calibration files, designed to be used by experienced scientists to conduct higher-level data processing. To benefit fully of the Swift data, non-experts in the field would have to go through a learning process that most people cannot afford. Even professionals may have significant problems if they have no direct experience with X-ray data analysis or need to analyse large amounts of data, like hundreds of observations. 

Following the directives of data transparency described in the previous section, the work presented here builds on the public domain data provided by Swift to deliver science-ready, high-level products for the instrument's entire blazar catalogue, increasing its usability to a larger number of potential users. This work is meant to be a concrete example of how the Open Universe Initiative can contribute to create effective solutions to data accessibility. Such efforts in turn hold a potential impact for research in various sectors of astrophysics and space science, and in the development of new, more user-oriented applications.

%\textcolor{blue}{Comment by Ana and Osvaldo: we can take care of a similar catalogue for GRBs and neutrinos, what do you think?}
%\textcolor{red}{PG: certainly for GRBs. As for neutrinos I have been working with the IceCube people in Germany for a long time. We are currently writing a paper on all IceCube Tracks. I have to check with them but I guess some cooperation could be possible.}
%Within the Open Universe initiative we aim at making the Swift data completely transparent, that is science ready and usable with a single click...

\section{A master list of blazars}\label{blazarlist}

Most blazars known to date have been found in radio, X-ray, or \gr\ surveys, or through multi-frequency searches, with a discovery rate that has been steadily increasing over the last several years. For this reason no recent complete lists exist. To assemble a reasonably up-to-date comprehensive catalogue of blazars we have combined the BZCAT 5th edition \citep{Massaro2015}, the list of blazars in the Fermi 3LAC catalogue \citep{3LAC} and the recently released 3HSP list of high energy peaked and extreme blazars, which includes 2011 objects, and is expected to have a contamination fraction from non-blazar sources that is significantly less then 4\% \citep{3HSP}. 

The resulting master list\foothref{http://openuniverse.asi.it/OU4Blazars/MasterList/} includes 5340 distinct blazars, 3561 of which are from BZCAT, 1353 are sources in the 3HSP catalogue that are not included in BZCAT, and the remaining objects are 3LAC blazars that are not listed in the BZCAT or the 3HSP catalogue.
The number of blazars in the master list might seem large since only 20 years ago just a few hundred blazars were known. However, the comparison with the almost 2 million catalogued QSOs \citep[see the million quasars catalogue V5.7;][]{Flesch2015}\foothref{https://heasarc.gsfc.nasa.gov/w3browse/all/milliquas.html} gives a clear measure of how rare blazars are compared to other AGN. Despite that, blazars are almost the only type of extragalactic source detected so far in the high energy \gr\ band, reflecting their highly energetic and extreme nature.

To determine how many blazars have been observed by Swift during its first 14 years of orbital operations, we have cross-matched our master catalogue of blazars with the list of all Swift-XRT observations carried out before 31 December 2018, using a matching radius of 12 arc-minutes, approximately the size of the XRT field-of-view. 
About 50\% of the blazars in our master list (2633 objects) are within the field-of-view of at least one XRT observation. If we consider only pointings carried out in PC mode with exposure of at least 200 seconds, this number reduces to 2585.

%U.B. At least in the last version of the code I saw, the master list was "frozen" to a given past date. Will the final version released with the paper have an automatic update of the master list implemented?

While most blazars have been observed just once or a few times, several have been observed repeatedly, up to over 1,000 times for the case of MRK 421 (in PC or WT mode), for a total of approximately 18,000 XRT exposures in PC mode. For technical reasons, Swift observations of a given target are frequently split into short exposures (corresponding to single OBS\_ID, typically of 1~ks each, see fig. \ref{fig:expo_distr}) repeated over a period of a few days, until the requested exposure time is reached. In an effort to represent the originally requested exposure times, typically between 3 and 10 ks, and to increase sensitivity, rather than single snapshots or observation IDs we have merged into a single "observation" all the exposures carried out within one week of each other. This merging procedure reduced to 11,399 the number of observations of blazars to be processed as described the next paragraph.

%{\color{red} U.B. About the operations of merging the short exposure data sets: It is not clear from the text if these merged, final data-sets, will be presented as such for the final user, as "one-week integration products", or if this is only an intermediate step of the processing. Will the user have the possibility to decide the temporal resolution of the data it can retrieve from the pipeline, $>$ multiples of 1 week? Will the same temporal resolution be available for all products. This is not clear from the next sections either.}

\begin{figure}[h]
\centering
\includegraphics[width=9.0cm]{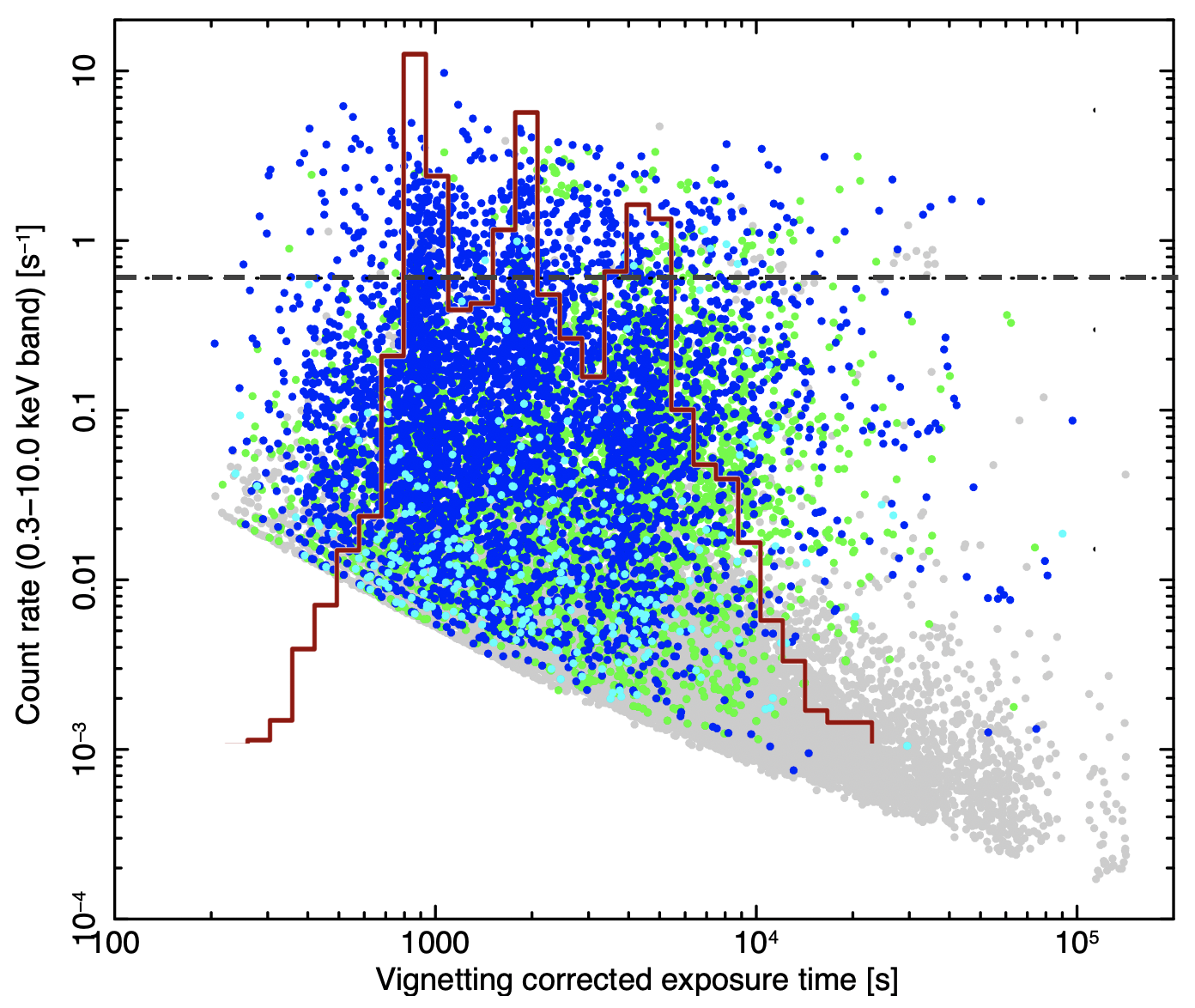}
\caption{Plot of the count rate in the 0.3-10.0 keV band versus the vignetting-corrected exposure time for all the sources detected in the 
data-set considered. FSRQs: blue points ; BL Lacs: green points; other targets and serendipitous sources: light grey points. The dotted horizontal line 
marks the count-rate limit above which pileup problems become significant in data taken in PC mode. The red histogram represents the number of blazars in each time bin, showing three peaks at approximately 1, 2 and 4-5ks, the preferred exposure times for blazars.}
\label{fig:ExpoVsCtr}       
\end{figure}

%\section{The \swiftdeepsky\ pipeline}\label{pipeline}

\section{Data processing}\label{processing}

To perform a detailed Swift-XRT image analysis of the blazars in the sample described above, we have developed a software pipeline called  \swiftdeepsky\, \citep{deepsky} building on the Swift X-Ray Telescope Data Analysis Software (XRTDAS), designed and developed under the responsibility of the Space Science Data Center (SSDC) of the Italian Space Agency (ASI) and distributed within the NASA HEASoft\foothref{https://heasarc.gsfc.nasa.gov/docs/software/lheasoft/} software package and related XRT calibration files (CALDB)\foothref{https://heasarc.gsfc.nasa.gov/docs/heasarc/caldb/}.
For a description of the \swiftdeepsky\, software see Appendix \ref{appendix}.

The processing of all the 11,399 one-week co-added Swift XRT PC mode observations of the blazars described in Section \ref{blazarlist} led to the detections of approximately 51,000 X-ray sources, including blazars (both observed as targets of Swift observations and as serendipitous sources), other Swift targets, and a large number of serendipitous sources. 

Some of the blazars observed by Swift are located in complex regions of the X-ray sky, where significant extended X-ray emission is present. These might be fields including bright clusters of galaxies or which suffer from residual contamination due to ``bright Earth'', that is, radiation reflected by the atmosphere when Swift points close to the Earth limb. Such fields can be very complex and certainly difficult to analyse in an automatic way or by inexperienced users. 
%\textcolor{blue}{Matteo: give also the number of observations filtered out by this cleaning procedure.} 
To remove this problem we applied a cleaning procedure that consisted of two checks:
\begin{enumerate}
\item \textbf{Background intensity consistent with expectations.} All fields where the measured background was found to be larger than 4$\sigma$ compared to the average image background observed in all fields in the sample were flagged out. 
\item \textbf{Number of detected point sources consistent with X-ray LogN-LogS.}
To do that we built a 80x80 pixels sensitivity map based on local exposure, vignetting, and background. We then used this map and the LogN-LogS of \cite{xmmlogns} to calculate the expected number of cosmic extragalactic sources in the field. All X-ray images where the number of detected sources deviated more than 4$\sigma$ from the expectations were flagged out. 
\end{enumerate}
This procedure lead to the identification of 353 observations ($\sim 3$\% of the total) that suffer from significant background or apparent point-source excesses. All sources detected in these fields, including 74 blazars, have been removed from the final sample. 

%{\color{red}Describe cleaning process here}
%{\color{magenta} comment by Sara T.: how a non-expert using the Docker version can be aware of the fact the data he/she requested should be flagged out? Is the Docker pipeline flagging them?}
%Answer to Sara by P.G. The flagging is done at the level of database ingestion, not in the Docker container. We are working with Carlos to include the cleaning in Docker.

This process led to a ``clean'' sample including a total of 33,396 X-ray sources, with 8,896 detections of 2,308 distinct blazars comprising 770 FSRQs, 1,276 BL Lacs and 262 blazars of uncertain type. The remaining 24,500 detection in the sample are serendipitous X-ray sources (mostly radio quiet AGN) or targets of XRT observations where one of the blazars in our reference list was detected as a serendipitous source.

Fig. \ref{fig:ExpoVsCtr} plots the observed count rate in the 0.3-10.0 keV band against exposure time of all the sources in the clean sample showing
FSRQs in blue, BL Lacs in green and other targets and serendipitous sources in grey. The dotted horizontal line marks the count-rate limit above which pileup \foothref{https://swift.gsfc.nasa.gov/analysis/xrt\_swguide\_v1\_2.pdf} becomes significant in sources observed in PC mode, causing significant flux underestimation and spectral distortion. 
%Detections with count rates in excess of 0.6 cts/s are excluded from the sample described above, which now includes 7,940 observations of 2,275 distinct blazars. 
%{\color{green} comments by Narek: 1) I remember that the count rate for pileup was 0.5 cts/s why we used 0.6 ?. 2) there are methods to deal with observations when count rate is $>$ 0.5 cts/s, in the future the script will allow to analyse such observations or not}.
%{\color{red} At ASDC we have always used 0.6 cts/s, it doesn't change much. We already take into account, and correct properly for pileup even for WT mode (MKN421 at its brightest needs that), in the other script that generates full spectra using XSPEC (see the SED of OJ 287 below). We will publish these spectra in another paper soon.} 

As can be seen from Fig. \ref{fig:ExpoVsCtr} most blazars are detected well above the minimum detectable count rate, and therefore with good statistics, as in most cases they were the target of the Swift observation. The red histogram overlaid on the plot represents the number of blazars included in each time bin. Three peaks at approximately 1, 2 and 4-5 ks are clearly visible, reflecting the Swift preferred exposure times for blazars.

The following scientific data products that match the operational definition of transparency given in Par. \ref{intro}, are available.
%{\color{red} The operational definition of "transparency" is not present anymore in the introduction as such - so, this needs to be re-written.}

\begin{enumerate}

\item A catalogue of XRT point sources called 1OUSXB, including the following 
\begin{itemize}
\item Count rates and X-ray fluxes in the 0.3-10.0 keV (full band), 0.3-1.0 keV (soft band), 1.0-2.0 keV (medium band), and the 2.0-10.0 keV (hard band).
\item Three sigma upper limits when a source is not detected in the soft, medium or hard energy band.
\item Three independent \nufnu\  flux measurements, de-absorbed to correct for Galactic absorption and suitable for SED plotting, calculated at 0.5, 1.5 and 4.5 keV, from the count rates observed in the soft, medium and hard energy bands. These \nufnu\ fluxes are available within the VOU-Blazars \citep{VOU-Blazars} and the SSDC-SED\footnote{\url{https://tools.ssdc.asi.it/SED/}} tools.
\item Power-law energy slope and one sigma statistical error, estimated from the hardness ratio, defined as (soft+medium band)/(hard band), only for sources detected with at least $\approx$ 50 net counts), and from a least-squares fit to the \nufnu\ fluxes at 0.5, 1.5 and 4.5 keV.
\item A 3.0 keV \nufnu\ flux point based on the count rate in the full band.
\item A \nufnu\ flux at 1.0 keV, interpolated between the fluxes at 0.5 and 1.5 keV, suitable for light curve comparison with other X-ray satellites at the reference energy of 1.0 keV. 
\end{itemize}

\item X-ray images in the full, soft, medium and hard energy bands. These images are provided in GIF format for easy inspection.

\item Co-added events files and associated-vignetting corrected exposure maps.

\item A ready-to-use Docker version of the \swiftdeepsky\ pipeline software, as well as the open source version.

\end{enumerate}

The results and data products are available in a variety of ways: (i) as FITS and CSV formatted tables; (ii) as on-line interactive tables that, in addition to the parameters included in the FITS catalogue, give access to products such as images; (iii) as SED spectral data accessible via the VOU-SED and the SSDC SED tools, all of which are available from the http://openuniverse.asi.it portal, (iv) through a
web query interface at vo.bsdc.icranet.org, and via Virtual Observatory services. (see http://openuniverse.asi.it/OU4Blazars for details). 
%{\color{green} comments by Narek: for the modeling of multi-wavelength SEDs, the X-ray flux points (6 or even higher) are very helpful, the script will provide such data ?}
%{\color{red} P.G. : all the SED data can be downloaded from both the VOU-SED and SSDC tools} 
The table is also published through an IVOA cone search service (SCS\foothref{http://www.ivoa.net/documents/latest/ConeSearch.html}) and database graphical query interface provided by DaCHS \citep{dachs} at
\href{http://vo.bsdc.icranet.org/ousxb/q/cone/form?__nevow_form__=genForm&MAXREC=10&_FORMAT=HTML&submit=Go}{\color{blue}{http://vo.bsdc.icranet.org/ousxb/q/cone/form}}.

In addition, full radio to \gr\ SEDs for each blazar, including the data of the results presented here, can be generated from the Open Universe web portal at \href{http://www.openuniverse.asi.it}{\color{blue}{openuniverse.asi.it}}. 

%{\color{red} U.B. It is not clear, for all of the products, how they are accessible? Or at least, the various resources are a bit confusing in the way they are listed, because they are not presented anywhere in detail. Since OU aims for transparency and findability, we should make this more clear here. An alternative could be to add an appendix about accessing the OU products of blazars, through the various services and portals, and in so doing, quickly explain the services the Initiative already built.} 

\subsection{Blazars frequently monitored by Swift}

A few bright and highly variable objects have been monitored by Swift-XRT, particularly during flares, with short observations performed on a daily basis. To avoid losing possible daily X-ray variability that was the objective of the frequent rate of observations for these sources we have run the \swiftdeepsky\ pipeline using a time interval of 1 day. 
The list of these intensely observed blazars and the total number of observations for each object is given in Table \ref{frequentblazars}, where
column 1 is the source name;
column 2 is the number of XRT PC mode observations in one day with exposure larger than 200 seconds;
column 3 is the number of detections free of pileup;
column 4 is the 0.3-10.0 keV median count rate; and
column 5 is the 0.3-10.0 keV median absolute deviation about the median as a variability measure.

All the data products generated for the observations integrated over one week are also available for these shorter exposures.

%{\color{red} Any differences on the products available for these sources with this "special" resolution? Spectra, image, SED points... And where they can be retrieved from, w.r.t to the other blazars' products?}

\begin{table*}
%\begin{small}
\begin{center}
\caption{Blazars observed by Swift with short and frequent, mostly daily XRT exposures in PC mode.}
\begin{tabular}{lcccc}
\hline\hline
Blazar name & no. of XRT   & no. of detections & $\langle\mathrm {rate}\rangle^b$ & MAD$^c$ \\
 &  observations$^a$ & below pileup limit & ($\mathrm{cts}~\mathrm{s}^{-1}$) & (\%)\\
\hline
OJ 287 & 304& 303 & 0.202 & 23.3 \\
3C 279 & 282 & 261 & 0.355 & 17.5 \\
PKS 1510-089 & 218 & 218 & 0.217 & 16.6 \\
S5 0716+714 & 185 & 156 & 0.294 & 34.0 \\
PG 1553+113 & 165  & 33 & 1.040 & 41.3\\
PKS 0235+164  & 158 & 155 & 0.048 & 40.5 \\
PKS 0208-512 & 140 & 140 & 0.063 & 23.3\\
BZQJ 1635+3808 & 130 & 130 & 0.105 & 47.1 \\
1ES 2344+514 & 126 & 64 & 0.599 & 26.7 \\
PKS 0528+134  & 116 & 116 & 0.035 & 48.0 \\
ON 231       & 113 & 112 & 0.057 & 25.8 \\
CTA 102     & 110 & 89 & 0.322 & 50.9 \\
1H 0323+342 & 108 & 88 & 0.426 & 23.7 \\
MS 1207.9+3945 & 103 & 103 & 0.084 & 21.5 \\
PKS 0921-213 & 94 & 94 & 0.216 & 19.9 \\
NRAO 530 & 89 & 89 & 0.045 & 19.0 \\
3C 66A  & 83 & 83 & 0.100 & 28.1 \\ 
PKS 1222+216 & 73 & 73 & 0.111 & 22.2 \\
TXS 0506+056$^{d}$ & 49 & 49 & 0.053 & 23.5 \\
\hline\hline
\end{tabular}
\label{frequentblazars}
\end{center}
$^a$ {\footnotesize Integrated with one day time interval executed in PC mode with exposure time larger than 200 seconds} \\
$^b$ {\footnotesize Median count rate in the total $0.3-10$ keV band} \\
$^c$ {\footnotesize Median Absolute Deviation about the median count rate} \\
$^d$ {\footnotesize Blazar associated to the IceCube high-energy neutrino IC170922 
%{\color{magenta} Comment by Sara T.: maybe add reference} 
and other IceCube neutrinos \citep{Dissecting}.}
\end{table*}
Some well-known blazars such as MRK~421, MRK~501 and 3C~454.3 that have been monitored by Swift do not appear in table \ref{frequentblazars} since their X-ray fluxes are so high that the Swift-XRT observations were mostly made in WT mode; even when PC mode observations are available in the archive, these are very often subject to serious pileup. \\
Detailed flux measurements, spectra and other high-transparency data products for these blazars will be published in a future paper presenting a systematic spectral analysis of WT and PC Swift XRT data of blazars \citep[see][for preliminary results]{Swift10years} using the XSPEC package \citep{xspec}.  
\section{Results: some examples}

In this section we give some examples of the content of our catalogue and related data, and of how it can be used for multi-frequency, time variability, and multi-messenger analysis.
%{\color{red} U.B. How it can be used for some specific science application, such as the multi-messenger science, or how it can be used in terms of tools...? Please, specify.}

\subsection{Sensitivity range}

\begin{figure}[h]
\centering
\includegraphics[width=9.0cm]{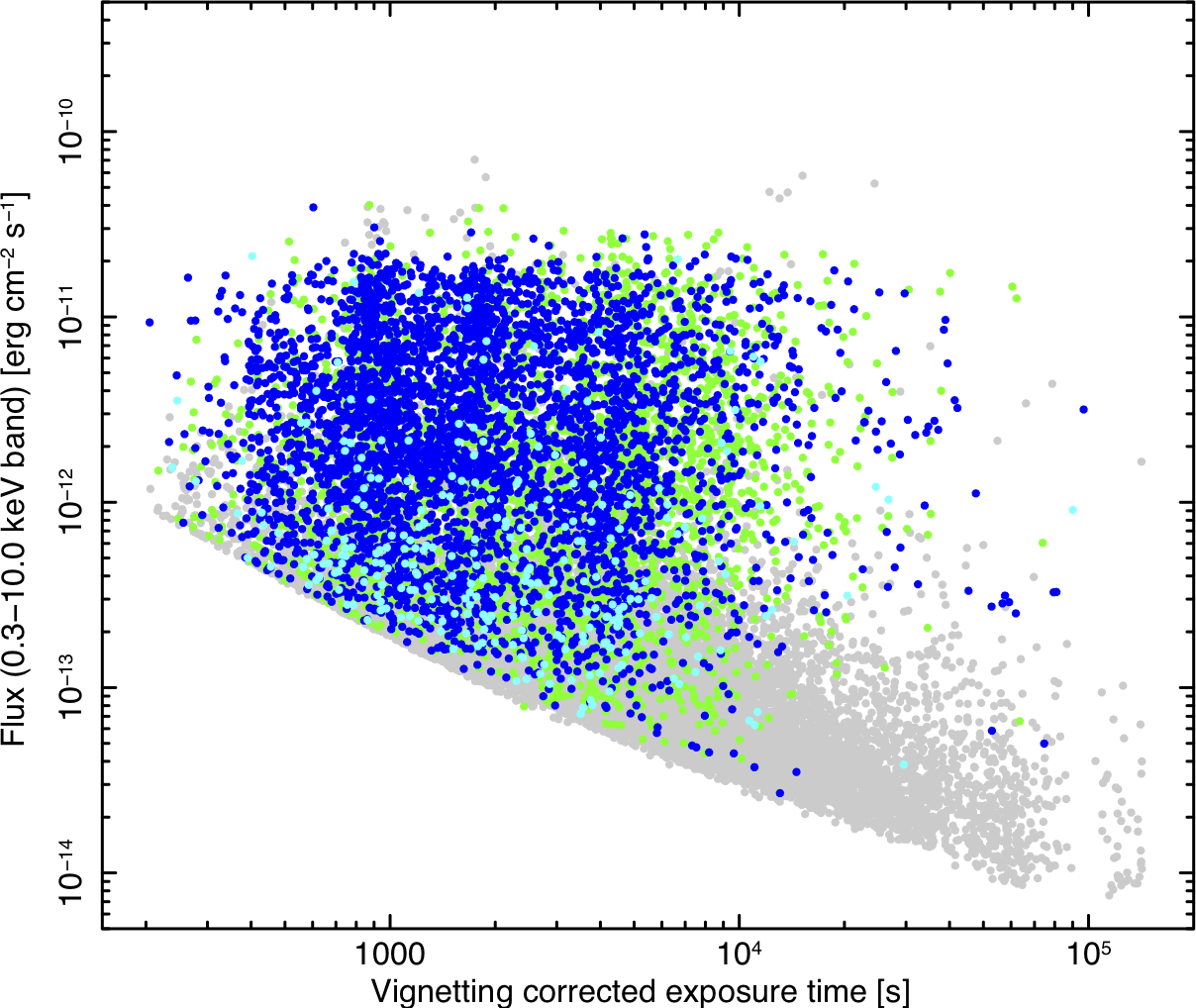}
\caption{0.3-10.0 keV band flux versus vignetting-corrected exposure time for all detected sources free of pileup in PC mode (count rate$<$ 0.6 cts/s). Colour coding as in \protect{Fig. \ref{fig:ExpoVsCtr}}.} 
\label{fig:ExpoVsFlux}       
\end{figure}

Fig. \ref{fig:ExpoVsFlux} plots the observed 0.3-10.0 keV energy flux vs exposure time, showing that the sensitivity of the Swift XRT data set considered ranges from $\sim 10^{-12}$ \ergs\ for the very shortest ($\sim$ 200 seconds) exposures to $\sim 7\times10^{-15}$\ergs\ for exposures as long as  $\sim10^5$ seconds. Sources detected with a 0.3-10.0 keV count rate larger than 0.6 cts/s are affected by pileup and are not plotted in this figure. For this reason the maximum flux in the sample of sources not flagged as affected by pileup is limited to $\sim 2-4\times10^{-11}$ \ergs, depending on the spectral slope and the amount of absorption in the Galaxy (NH). This range of sensitivities is intermediate between that of the sources in the catalogue of \cite{DElia2013}, who only considered single observation\_IDs, and the lists of \cite{Evans2014,Puccetti2011}, who instead stacked all the XRT data available. 

\subsection{Energy (SEDs) and time domain (light-curves) plots}

\begin{figure}[h]
\centering
\includegraphics[width=9.0cm]{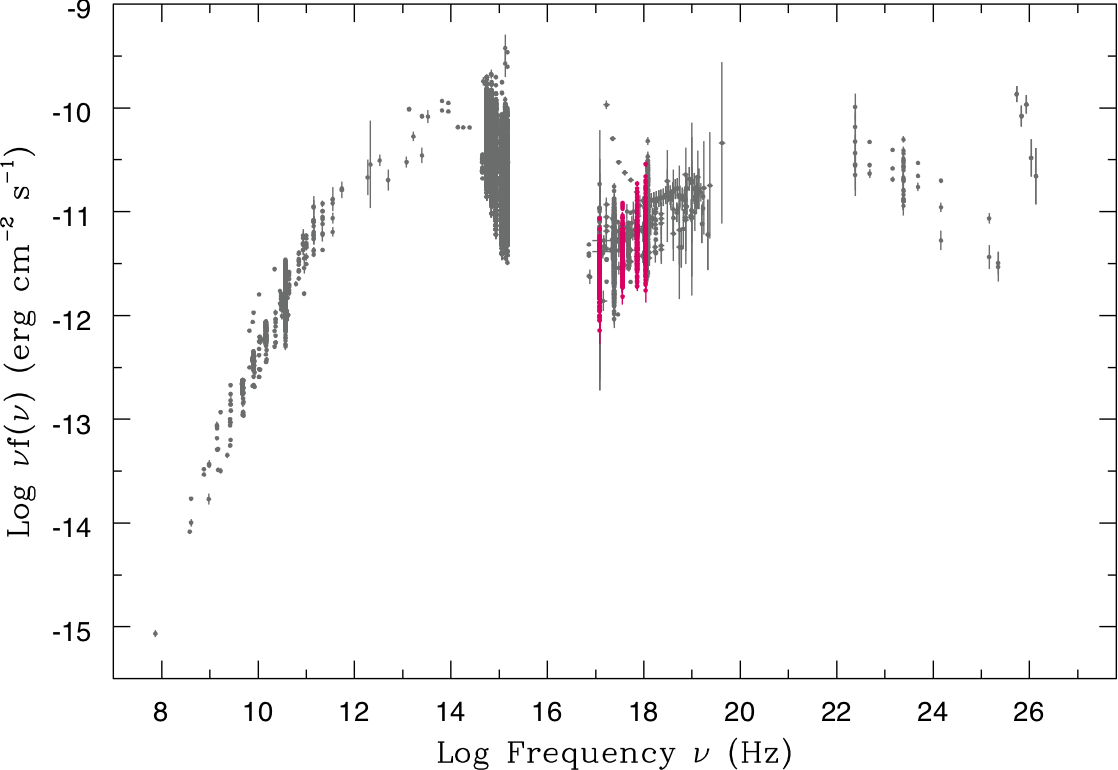}
\caption{Spectral Energy Distribution of BL Lacertae. Grey points are archival multi-frequency data. Red points are from this work.}
\label{fig:SED-BLLac}       
\end{figure}

\begin{figure}[h]
\centering
\includegraphics[width=9.0cm]{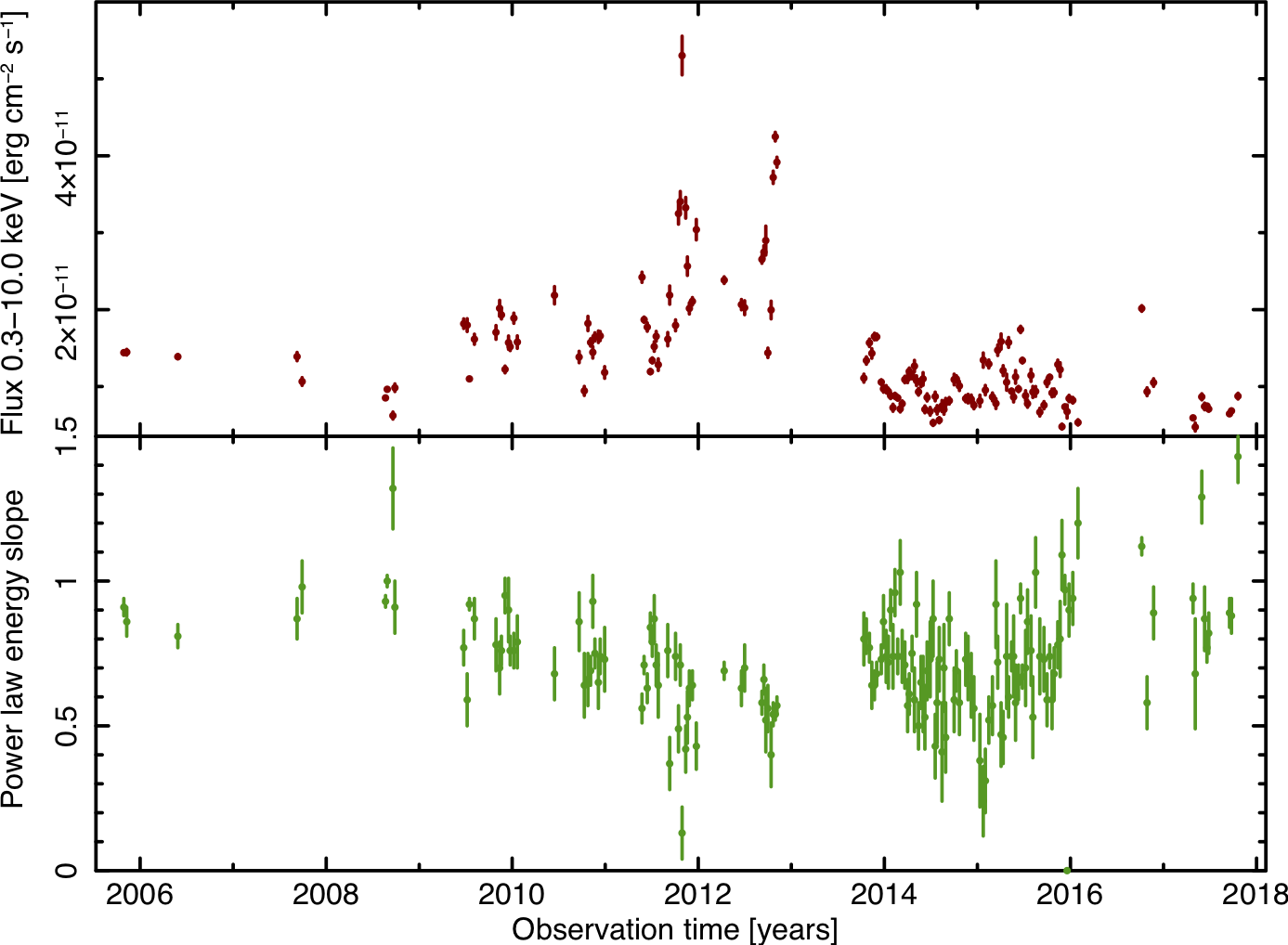}
\caption{History of the 0.3-10.0 keV band flux and estimated power-law energy spectral slope of BL Lacertae.
%{\color{red}TO BE REDONE using spectral slopes from the fitting of \nufnu\ data.}
}
\label{fig:LC-BLLac}       
\end{figure}

Figure \ref{fig:SED-BLLac} shows the SED of BL Lacertae, the prototypical object of the blazar class. The grey points are non-simultaneous multi-frequency archival data showing the typical double-humped SED and large flux variations. The red points are 0.5, 1.5, 3.0 and 4.5 keV \nufnu\ measurements from our catalogue. They show a hard spectrum and approximately a factor 10 intensity variability, the details of which are shown in Fig. \ref{fig:LC-BLLac} where the 0.3-10 keV X-ray flux (top panel) and the energy spectral slope (bottom panel), are plotted as a function of observation date.

\begin{figure}[h]
\centering
\includegraphics[width=9.0cm]{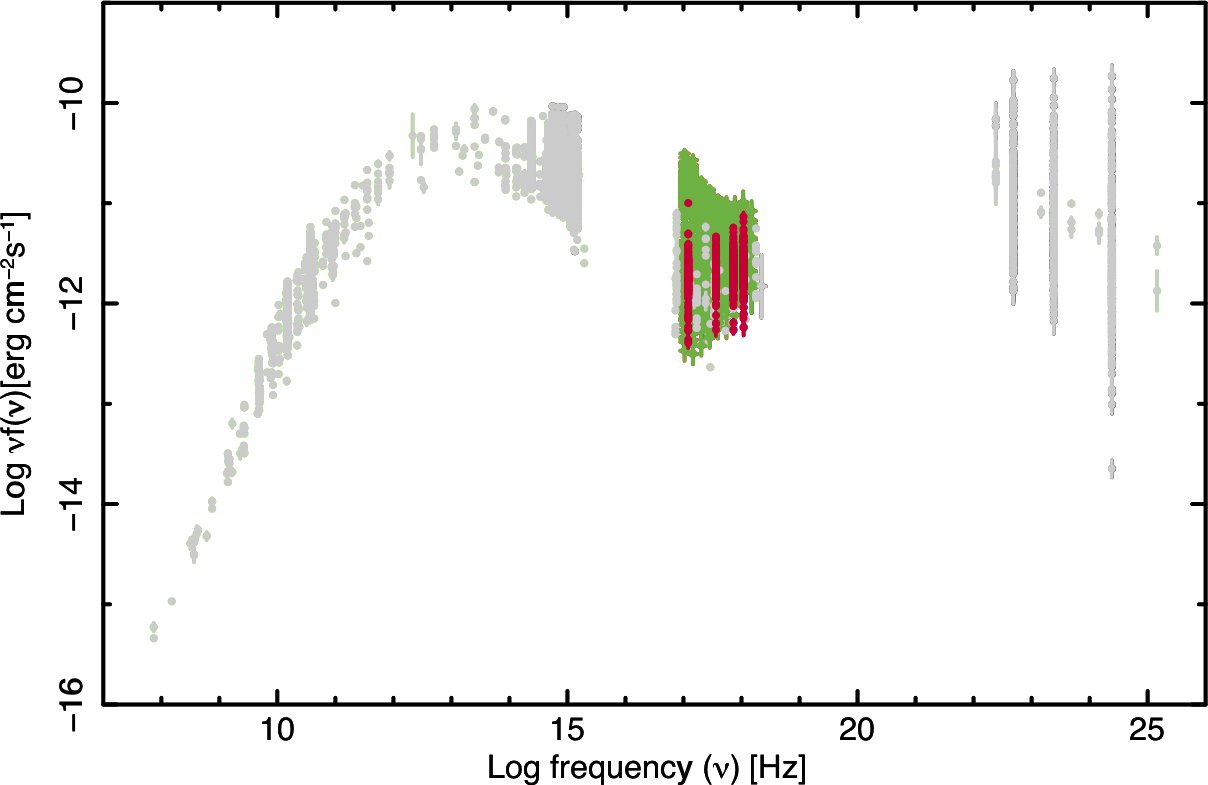}
\caption{Spectral Energy Distribution of the blazar OJ~287. Grey points are archival multi-frequency data. Red points are from this work (data in PC mode and count rate $<$0.6 cts/s). Green points are XSPEC spectra from the first analysis of all the XRT data in PC and WT mode run on a small sub-sample of blazars \citep{Swift10years}.}
\label{fig:SED-OJ287}       
\end{figure}

\begin{figure}[h]
\centering
\includegraphics[width=9.0cm]{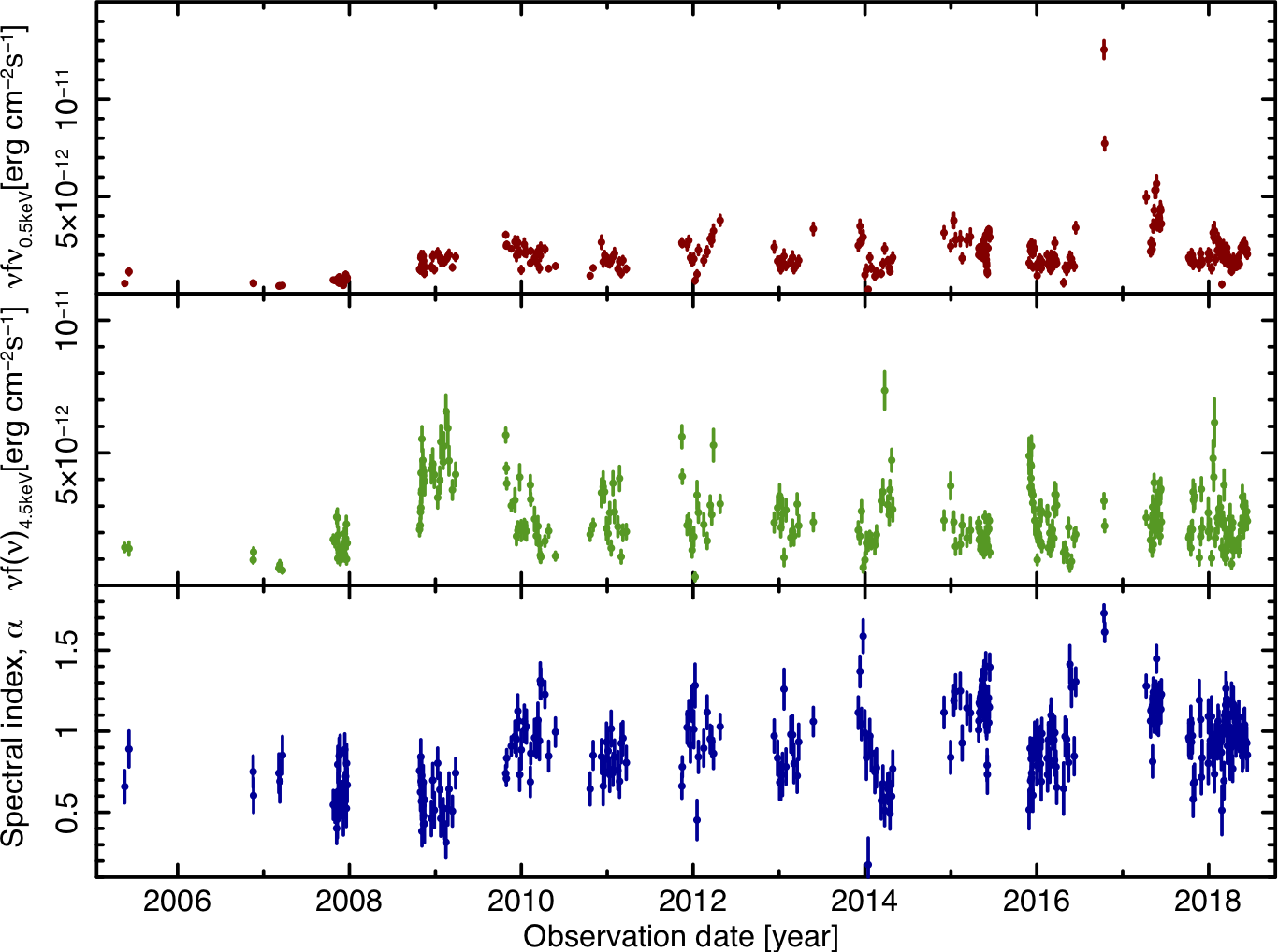}
\caption{The \nufnu\ flux at 0.5, 4.5 keV and the power law energy spectral index of the blazar OJ~287 are plotted against observation date. Many differences in the top and middle panel are clearly visible. Note in particular that the X-ray flares occur in different periods in the soft (0.5 keV) and hard (4.5 keV) part of the X-ray spectrum.
%\textcolor{blue}{This should be Figure 7 (it is cited in the text before the plot showing the light-curves of TXS 0506)}
}
\label{fig:LC-OJ287}       
\end{figure}

Fig. \ref{fig:SED-OJ287} 
%\textcolor{blue}{(We have still not mentioned Fig. 6 )}
shows the SED of OJ 287, a blazar where the two SED components meet in the middle of the XRT energy band. The green points are spectral data obtained from the first XSPEC spectral analysis of all the XRT data in PC and WT mode that was run on a sub-sample of bright blazars in 2015. 
%{\color{magenta} by Sara T.: are also these WT data to be release or are they release yet?}
%P.G. all the WT data are open, like all the swift data. They are not part of this release, but they are available within the VOU-Blazar/VOU-SED tool
The red points from this work do not reach the highest fluxes because of pileup problems above the count-rate of 0.6 ct/s. 
% the brightest observations were made in WT mode.
The top panel of figure \ref{fig:LC-OJ287} shows how the low energy flux (0.5 keV), at the end of the first SED component varies in a different way than the flux at 4.5 keV (middle panel), which belongs to the high energy SED component. The variation of the spectral slope is shown in the bottom panel.

%{\color{red}U.B. Can we say something more about these examples, that stress the uniqueness of what is done here? Like the combination of PC and WT data in the catalogue releases, if this is the case, etc.? The section is merely illustrative as it is now. It would further improve this illustrative character if we could stress with which specific resources and easiness the results were generated, and with which quality and quantity of information, thanks to the data products of the initiative.}

\subsection{The blazar TXS~0506+056: a likely source of IceCube neutrinos}

One way to illustrate the reliability of our database is to compare our results to those published by independent authors on the same data set. We consider here the case of TXS~0506+056, the blazar that has been recently, and for the first time, associated to several IceCube neutrinos, and in particular to the high-energy neutrino IC170922. This source was observed by Swift a few hours after the neutrino arrival, and has been frequently monitored in the weeks after the IceCube event.

\begin{figure}[h]
\centering
\includegraphics[width=9.0cm]{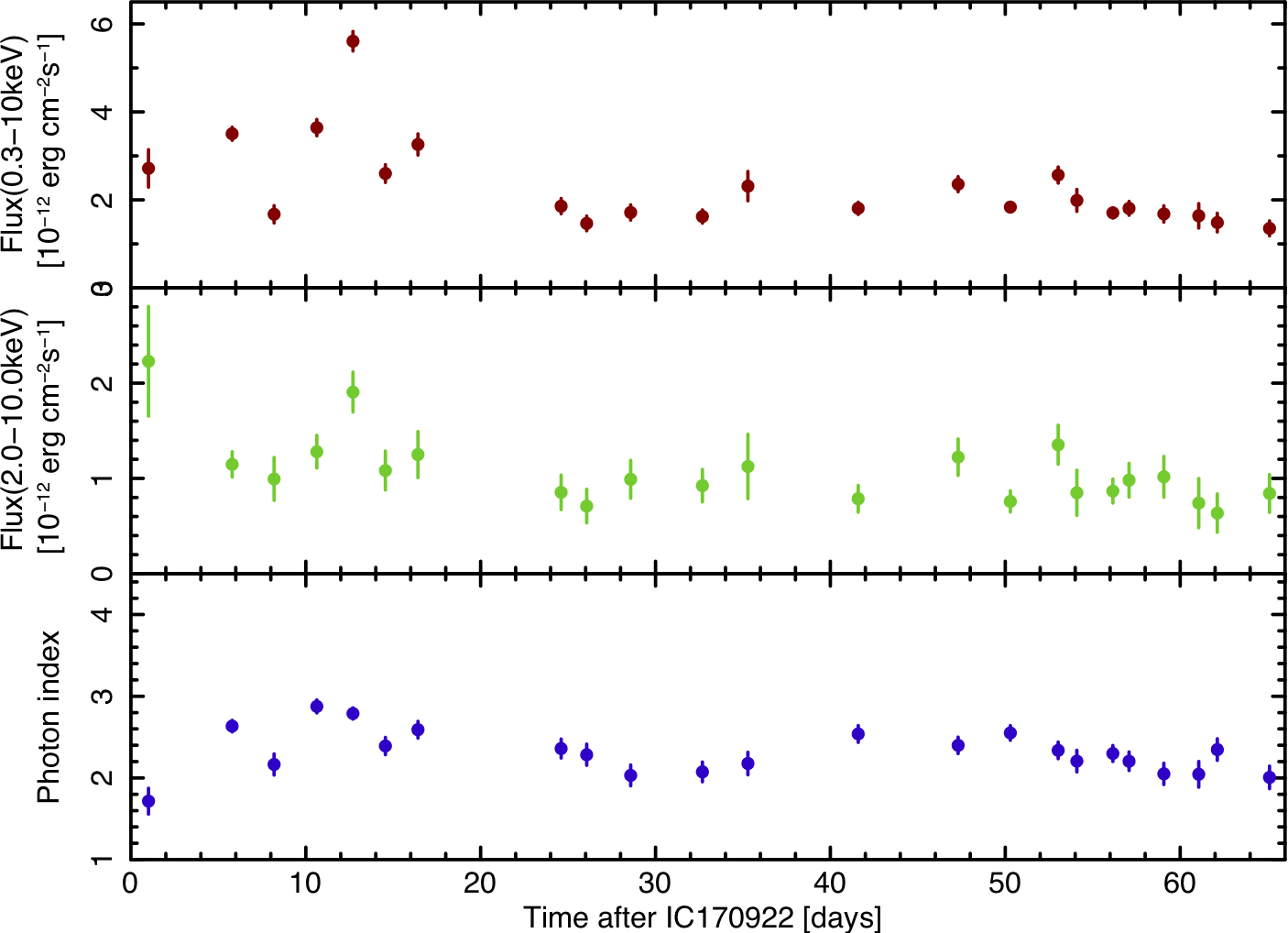}
\caption{The 0.3-10.0 keV and 2-10 keV band flux in units of 10$^{-12}$\ergs  (top and middle panel), and the power law photon index of TXS0506+056 (bottom panel) are plotted as a function of time. All uncertainties are one sigma.
%\textcolor{blue}{This should be Figure 8}
}
\label{fig:keivani}       
\end{figure}

A complete analysis of the first two months of the Swift-XRT monitoring of TXS~0506+056 has been reported by \cite{keivani}. Fig. \ref{fig:keivani} plots the 0.3-10.0 keV and 2-10 keV bands, and the power-law photon index of this blazar as a function of time, based on the content of our database which, we recall, was built in a completely automated way and could have been produced by any user with the \swiftdeepsky\, Docker container. A detailed comparison of this figure with Fig. 2 and Table 1 of \cite{keivani} clearly shows that our results are essentially identical to those of \cite{keivani}.
%{\color{red} U.B. Could we superpose those here to facilitate the comparison to the reader?} 
%PG, our results are essentially identical to those of Keivani et al. 
% if we tried to superpose one on top of the other we would only see one
%data set
Note that Fig. \ref{fig:keivani}, in addition to the flux in the full band (red points in the upper panel) reported by \cite{keivani}, also plots as green points in the middle panel the flux in the hard 2-10 keV band. We have done this to show that the maximum source intensity in the two energy bands occurs at different times: immediately after the arrival of the neutrino in the hard band, when the spectrum was hardest, and at a later time in the full 0.3-10.0 keV band.

\subsection{Some population statistical properties}
%{\color{red} I find this could come before the example applications, because it is more general.}
%Done 

%{\color{red} How does it compare to the sensitivity of the previous point-source catalogues cited?}

\begin{figure}[h]
\centering
\includegraphics[width=9.0cm]{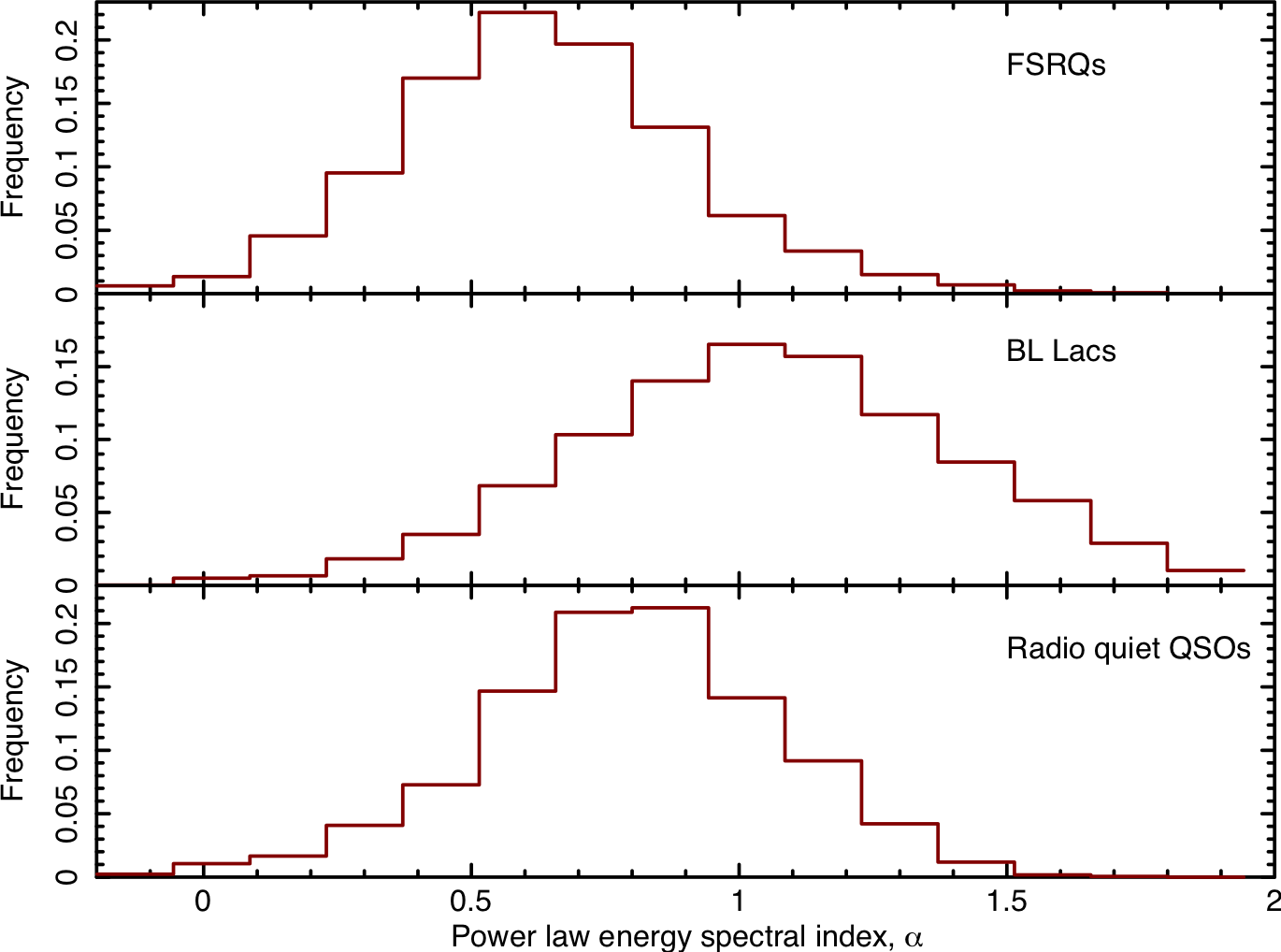}
\caption{The distribution of the power law energy spectral index in the sub-samples of FSRQs (top), BL Lacs (middle), and radio quiet QSOs (bottom).}
\label{fig:alphadistr}     
\end{figure}

Fig. \ref{fig:alphadistr} shows the distributions of the power-law energy spectral indices in the sub-samples of FSRQs, BL Lacs and QSOs with no radio detection in the million quasar catalogue \citep[MILLIQUAS][]{Flesch2015}, that have been serendipitously detected in our X-ray images. The distributions are clearly distinct, reflecting the different ranges of synchrotron peak energies of the SED components in FSRQs and BL Lacs \citep[see e.g.][]{GiommiPlanck,Giommi2012} and the different X-ray emission process in radio-quiet QSOs that is due to accretion onto the super-massive black hole, rather than non-thermal emission from a relativistic jet. The spectral index distribution in sample of QSOs with no radio emission peaks at $\sim$0.8, as expected in this type of sources \citep[e.g.][]{Comastri}.
%{\color{red} U.B. Can something more be added about the relevance of the result in comparison to previous similar samples in the literature?}

\begin{figure}[h]
\centering
\includegraphics[width=9.0cm]{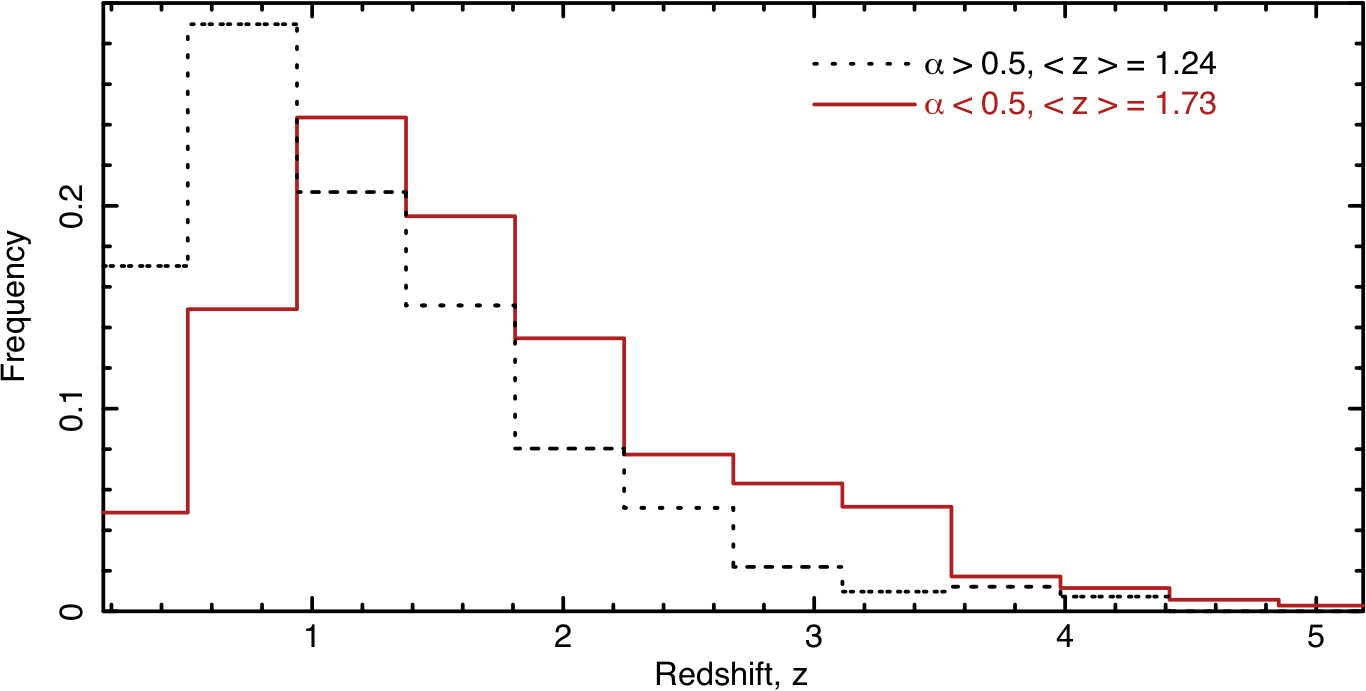}
\caption{The redshift distribution of FSRQs with different power law  spectral indices ($\alpha$). The black dotted histogram refers to objects with $\alpha > 0.5$ while the red solid histogram is for sources with  $\alpha < 0.5$. FSRQs with flat spectral slopes tend to have larger redshifts.}
\label{fig:redalpha}     
\end{figure}

Finally, Fig. \ref{fig:redalpha} illustrates the difference between the redshift distributions of FSRQs with hard ($\alpha < 0.5$, solid red histogram) and with steeper ($\alpha > 0.5$, dotted black histogram) 
X-ray spectral slopes. 
%\textcolor{blue}{A.V.P.: shouldn't it be: ...hard ($\alpha > 0.5$, dotted black histogram) 
%and with softer ($\alpha > 0.5$, ) X-ray spectral slopes. ? 
A KS test gives a probability of $\sim 2.\times10^{-7}$ that the two distributions come from the same parent population. The fact that the hard sources are statistically observed at higher redshifts could be due to different reasons, e.g.: 
\begin{itemize}
    \item a spectral hardening at energies above 10~keV that enters the XRT energy band (0.3-10.0~keV) only for high-redshift sources.
    \item a dependence of the X-ray spectral slope on luminosity, with the high-redshift and more powerful FSRQs showing the hardest spectra.
    \item the X-ray spectrum is the result of the superposition of two components, one that is relatively soft ($\alpha \sim 0.8$) due to accretion onto the super-massive black hole, and a second, much harder one that is due to the inverse-Compton emission from the jet. At relatively high energies (E $\gsim 10$~keV) the hard component starts dominating the spectrum, which is red-shifted into the 0.3-10~keV XRT band for high redshift sources.
%    \item a selection effect where powerful high Compton dominance sources 
\end{itemize}
A detailed statistical study is needed to distinguish among the various possibilities.

\begin{table*} [h!]
\begin{tiny}
\begin{center}
\caption{Details of the content of the 1OUSXB table in FITS format}
\begin{tabular}{llll}
\hline\hline
Column & Format & Units & Description\\
\hline
Blazar name    & 18A & & Blazar name as it appears in Table \ref{frequentblazars} \\
 & & & or in the 5BZCAT, 3LAC or 3HSP catalogue\\
Source name & 22A & & catalogue source name formatted as 1OUSXBJhhmm.f+/-ddmm \\
 & & & hhmm=hours, min of R.A., f=fraction of minutes \\
 & & & ddmm=deg and min of Dec.  \\ 
Other\_name & 22A & & alternative source name from literature when available \\
 R.A.         & D & deg & Right Ascension in degrees (J2000.0 epoch) \\ 
 Dec.                 & D & deg & Declination in degrees (J2000.0 epoch) \\
 MJD\_Start      & D & days &  Modified Julian Day of observation start time  \\
 MJD\_End       & D & days &  Modified Julian Day of observation end time  \\
 Exposure\_time & D & s & Vignetting corrected exposure time$^{a}$ \\
 Ctr\_03\_10     & D & cts s$^{-1}$ & Count rate in the full (0.3-10.0 keV) band\\
 Ctr\_03\_10\_error   & D & cts s$^{-1}$ &  1 sigma error on Ctr\_03\_10 \\
 Ctr\_03\_1       & D & cts s$^{-1}$& Count rate in the soft (0.3-1.0 keV) band \\
  Ctr\_03\_1\_error & D & cts s$^{-1}$ & 1 sigma error on Ctr\_03\_1 \\
  Ctr\_03\_1\_UL & D & cts s$^{-1}$ & 3 sigma upper limit$^{b}$ on Ctr\_03\_1 \\
 Ctr\_1\_2         & D & cts s$^{-1}$ & Count rate in the medium (1.0-2.0) keV band\\
 Ctr\_1\_2\_error &D & cts s$^{-1}$ & 1 sigma error on Ctr\_1\_2 \\
 Ctr\_1\_2\_UL & D & cts s$^{-1}$ & 3 sigma upper limit$^{b}$ on Ctr\_1\_2 \\
 Ctr\_2\_10      & D & cts s$^{-1}$& Count rate in the hard (2.0-10.0 keV) band \\
 Ctr\_2\_10\_error & D & cts s$^{-1}$ & 1 sigma error on Ctr\_2\_10 \\
 Ctr\_2\_10\_UL & D & cts s$^{-1}$ & 3 sigma upper limit$^{b}$ on Ctr\_2\_10 \\
 Flux\_03\_10   & D &  erg cm$^{-2}$ s$^{-1}$ & Observed energy flux in the full band \\
 Flux\_03\_10\_error & D & erg cm$^{-2}$ s$^{-1}$& 1 sigma error on Flux\_03\_10 \\
 Flux\_03\_1  & D &  erg cm$^{-2}$ s$^{-1}$ & Observed energy flux in the soft band \\
Flux\_03\_1\_error & D & erg cm$^{-2}$ s$^{-1}$ &  1 sigma error on Flux\_03\_1 \\
Flux\_03\_1\_UL & D & erg cm$^{-2}$ s$^{-1}$ & 3 sigma upper limit$^{b}$ on Flux\_03\_1 \\
 Flux\_1\_2       & D & erg cm$^{-2}$ s$^{-1}$ & Observed energy flux in the medium band \\
 Flux\_1\_2\_error & D & erg cm$^{-2}$ s$^{-1}$ & 1 sigma error on Flux\_1\_2 \\
 Flux\_1\_2\_UL & D & erg cm$^{-2}$ s$^{-1}$ & 3 sigma upper limit$^{b}$ on Flux\_1\_2 \\
 Flux\_2\_10    & D & erg cm$^{-2}$ s$^{-1}$ & Observed energy flux in the hard band\\
 Flux\_2\_10\_error & D & erg cm$^{-2}$ s$^{-1}$ &  1 sigma error on Flux\_2\_10 \\
 Flux\_2\_10\_UL & D & erg cm$^{-2}$ s$^{-1}$ & 3 sigma upper limit$^{b}$ on Flux\_2\_10 \\
 Nufnu\_3.0keV & D & erg cm$^{-2}$ s$^{-1}$ & absorption corrected$^{c}$ \nufnu\ flux at 3.0 keV calculated from Ctr\_03\_10 \\
 Nufnu\_3.0keV\_error  & D & erg cm$^{-2}$ s$^{-1}$ &  1 sigma error on  Nufnu\_3.0keV \\   
 Nufnu\_0.5keV & D & erg cm$^{-2}$ s$^{-1}$ & absorption corrected$^{c}$ \nufnu\ flux at 0.5 keV calculated from Ctr\_03\_1 \\
Nufnu\_0.5keV\_error & D & erg cm$^{-2}$ s$^{-1}$ & 1 sigma error or 3 sigma upper limit $^{d}$ on Nufnu\_0.5keV \\
 Nufnu\_1.5keV & D & erg cm$^{-2}$ s$^{-1}$ & absorption corrected$^{c}$ \nufnu\ flux at 1.5 keV calculated from Ctr\_1\_2  \\
Nufnu\_1.5keV\_error & D & erg cm$^{-2}$ s$^{-1}$ & 1 sigma error or 3 sigma upper limit $^{d}$ on Nufnu\_1.5keV \\
 Nufnu\_4.5keV & D & erg cm$^{-2}$ s$^{-1}$ & absorption corrected$^{c}$ \nufnu\ flux at 4.5 keV calculated from Ctr\_2\_10  \\
Nufnu\_4.5keV\_error & D & erg cm$^{-2}$ s$^{-1}$ & 1 sigma error or 3 sigma upper limit $^{d}$ on Nufnu\_4.5keV \\
 Nufnu\_1.0keV & D & erg cm$^{-2}$ s$^{-1}$ & \nufnu\ flux at 1.0 keV interpolated between Nufnu\_0.5keV and Nufnu\_1.5keV  \\
Nufnu\_1.0keV\_error & D & erg cm$^{-2}$ s$^{-1}$ & 1 sigma error on Nufnu\_1.0keV\\
Spectral\_slope & D & & Power law energy spectral index  \\
Spectral\_slope\_error & D & & 1 sigma error on Spectral\_slope \\
Pileup flag & 3A & & YES if Ctr\_03\_10 $> 0.6$ cts s$^{-1}$, NO otherwise \\
\hline\hline
\end{tabular}
\label{fitstable}
\end{center}
%\begin{tablesnote}
$^a$ {\footnotesize The vignetting-corrected exposure time is equivalent to the exposure with the source exactly at the centre of the field of view.}\\
$^b$ {\footnotesize Upper limits are provided when no significant detection is achieved in the energy band considered.}\\
$^c$ {\footnotesize Galactic-absorption corrected flux in the observer frame.}\\
$^d$ {\footnotesize In case the source is not detected in this band, the Nufnu flux is set to 0 and the error is set equal to a three sigma upper limit} 
%\end{tablesnote}
\end{tiny}
\end{table*}

\subsection{Comparison with previous Swift-XRT catalogues}

Other catalogues of point-like X-ray sources detected in Swift-XRT images have been published in the past, e.g. \cite{Puccetti2011,DElia2013,Evans2014}. In addition to the coverage extension from 8 to 14 years of Swift observations, the approach followed in this paper is different in several ways:

\begin{itemize}
\item It has been designed to make all scientific products easily and readily accessible, in a number of ways (e.g. through VO access, the Open Universe portal, via SED tools, etc.). 
\item It demonstrates the possibility to extend the access to X-ray data analysis, and space science data in general, to non-experts, in principle to anyone interested in the field.

%\item It is based on 14 years of Swift-XRT data, compared to at most 8 years of data of previous catalogues. %\textcolor{blue}{Suggestion: Since Open Universe catalogue covers the total observation time of Swift (14 years), it will improve the systematics compared to the existing catalogues.}

\item It intends to support present and future gamma-ray surveys, in the context of the emerging field of multi-messenger astrophysics, by providing suitable blazar data from the most extensive X-ray mission of its kind. Scientific applications of this dataset could include: 
%focuses on the use case of blazars, with the intention to support present and future \gr\ surveys and the opening field of multi-messenger astrophysics. This will, in particular: 
a) help in the identification of HE and VHE \gr\, sources, as well as  high-energy neutrinos and possibly UHECRs astrophysical counterparts; b) support the selection of targets for VHE \gr\ observations; c) provide multi-temporal X-ray data suitable for the construction of time dependent SEDs to be compared to physical models, etc. 

%{\color{magenta} by Sara T. maybe we should detail more why this work will support future \gr\ surveys} \textcolor{blue}{Suggestion: as we have discussed throughout the paper and exemplified with data analysis from blazars, the purpose of the Open Universe blazar catalogue is to facilitate the access to updated observation data and to process them by using full-proof instruments for analysis (software).}
% {\color{red} U.B. I agree with the comment by Sara here. Most emphasis throughout the text was put to the work as an exemplification of the Open Universe Initiative. Only once does the argument of the multi-messenger comes in. So the mention here is a bit abrupt. I think the point could be raised elsewhere in the text as well, in more detail, to contextualise it better, such as in the introduction or section 2. Also, why does this catalogue do a)-c) better than the previous point-source catalogues?}

% \item It can be updated at any moment by anyone. This is the most innovative aspect of our approach.

%{\color{red} These are the two driving forces of the work, such as presented - the title of the paper reads Open Universe for Blazars", so they should come before others in the list, I think, with the item on "comply with principles" coming first.} 

\end{itemize}

\section{Open Universe for blazars: future developments}\label{future}

We intend to extend the work presented in this paper to other multi-frequency data on blazars. The following is a preliminary list of activities planned to be carried out in the coming months.

The Swift XRT data of blazars with more than 50 net counts in our database (about 1,000 objects), as well as all XRT data of bright blazars taken in both WT readout mode will be processed by means of an extended pipeline that takes into account and corrects for pileup problems, and generates X-ray spectra using the XSPEC package \citep{xspec}.
Early results from a test run of this processing, carried out on a small subset of very bright X-ray blazars, have been reported in \cite{Swift10years}.
%Preliminary results of this processing, that corrects for pileup problems, have been presented in \cite{Swift10years}.

A similar procedure will be followed to analyse all public observations of blazars carried out with the hard X-ray observatory NuSTAR \citep{nustar}.

Transparent-level results (\nufnu\, spectra with 10 energy points) and high-level data products (pha spectra and response files) will also be made available in a number of ways including VOU-Blazars, the SSDC SED tool and other services available within the VO and the Open Universe portal.

As part of the Open Universe for blazars project, we are also producing, and will soon make available, Swift BAT spectra of the brightest X-ray blazars, and a set of adaptive-bin gamma-ray light-curves from Fermi-LAT data. These will be accessible, as with the other products, from the Open Universe portal, the VOU-Blazars and the SSDC SED tools, as well as via VO services. The temporal and spectral information will be combined to generate SED movies similar to the ones published in \cite{Dissecting}, and  \cite{Swift10years}, which are available on-line at https://www.youtube.com/embed/lFBciGIT0mE and http://youtu.be/nAZYcXcUGW8 \\

%As part of the Open Universe for blazars project we also plan to produce and make available on the openuniverse.asi.it portal, the VOU-BLazars and SSDC SED tools, as well as through VO services, Swift BAT spectra of the brightest X-ray blazars, and a set of adaptive-bin \gr\ light-curves \citep{Adaptive} using Fermi-LAT data \citep{Fermi-LAT}, and SED movies similar to the ones 

\subsection{An interactive database that is always up to date}\label{database}
%{\color{red} U.B. Alternative title suggestion: "An interactive database that is always up to date"}
One of the practical challenges in the provision of open data services in astronomy, especially when these involve operational satellites and aim at delivering high-level products in a timely manner, is to maintain the database up-to-date and conforming to the latest versions of the data analysis software and calibration files.
%{\sout Any astronomical catalogue or scientific database from operational satellites inevitably become outdated soon after their release as new observations are made and new versions of the data analysis software and calibration files are released.}
Despite the practical difficulties involved in achieving this, such standards constitute an important step towards improving data transparency and in democratising the accessibility to the best quality scientific data available.

%difficulties involved, the timeliness of the datasets and to assure that they comply with the highest standards of quality of the analysis software is fundamental to the objective of the Open Universe Initiative, of democratising data access to all potential users, with the minimum possible distinction of quality and extent of data products and its usability.
%{\sout In the spirit of the Open Universe Initiative, and} 

By taking advantage of new software technologies, it is possible today to minimise such shortcomings. In the current application of ``Open Universe for Blazars'' we have had recourse to the use of Linux containers based on Docker technology to test a new approach on data publication to enable users seamlessly to contribute to updating the public, worldwide available database of Swift observations. The effective goal is to have the table of Swift-DeepSky results always up-to-date with the latest observations taken by Swift. The container-based distributed analysis model means that the data processing and updating of the database will follow the actual user demand. %where the actual processing and extraction of new information would naturally follow user demand.

Such a service -- \textit{i.e.}, the software link between the pipeline and a central VO-enabled service equipped with an automated data publishing pipeline -- is currently under internal commissioning. In the months to come such integration should become public, it will have \emph{no} impact on the client side of \swiftdeepsky. The current automatic publication workflow under test will provide two versions of results table: the \emph{best-of-all sources} (boa) table where each source appear only once -- where the signal-to-noise ration \textit{and} total exposure-time are maximum --, and the \emph{best-on-times sources} (bot) table, where sources may appear more than once for different epochs.

\section{Conclusion}\label{conclusion}

We presented new sets of astronomical data products based on 14 years of Swift XRT observations of blazars. The associated results database includes flux measurements, spectral and timing information, as well as other high-level material that can be accessed through the Virtual Observatory, the Open Universe portal and in other ways.
The pipeline software used to generate these data is also available in the form of an easy-to-use Docker container, which automatically downloads the low-level data and calibration necessary for the analysis, and can be run on all modern operating systems (Linux, Mac, Windows), thus removing platform dependencies and the need of any X-ray data analysis expertise. 

%{\color{magenta} by Sara T. to avoid words repetition with previous sentence, I suggest to substitute This approach with Using containerization technology ... or something similar} 
Our approach significantly lowers technical barriers to X-ray data analysis, and introduces the possibility for anyone to reproduce the content of the database, extend it to other source types, or update it when new observations become available or when a new version of the software or calibration data are released. This is a major evolution compared to the traditional approach where astronomical catalogues are generated by teams of experts using complex hardware infrastructure and are issued at irregular time intervals, often separated by several years.  \\  

This paper advances, therefore, the first steps towards the implementation of a high-transparency, multi-frequency dynamic database service for the high-energy astrophysics community, which may become a particularly relevant resource in the emerging field of multi-messenger astronomy.
%can be considered a first step towards the implementation of a high-transparency multi-frequency dynamical database as a service to the community operating in high-energy astrophysics and in the emerging field of multi-messenger astronomy. 

Following the principles of the Open Universe Initiative, this work is also meant to serve as a pathfinder to software solutions aiming to improve the usability of existing open space science data for a broader community of non-experts (scientists or otherwise) at a marginal cost. This is of particular importance in the context of the United Nations Sustainable Development Goals, as easy of access and transparency are key factors in the conversion of data into knowledge, and to achieve equal opportunities in the access of space science data and scientific information in general. Likewise, it is fully aligned with the mission of the United Nations Office for Outer Space Affairs (UNOOSA) in promoting international cooperation on the peaceful uses of outer space, and with the mandate of the Office following the UNISPACE conferences.

%This work is also meant to serve as a pathfinder, a feasibility demonstrator for the possibility to bring the usability of existing open space science data to the level of non-experts at no or marginal cost, one of the crucial objectives of the Open Universe initiative. This is of particular importance as easy accessibility and transparency are key factors for the conversion of data into knowledge and for the achievement of equal opportunity in the access to space data and scientific information in general, a valid contribution to the achievement of the United Nations Sustainable Development Goals 
%\textcolor{magenta} {and it is fully aligned with the mission of the United Nations Office for Outer Space Affairs in promoting international cooperation on the peaceful uses of outer space and the mandate of the Office following the UNISPACE conferences.
%}

We encourage researchers, students, or anyone interested in X-ray astrophysics, to use this method and our software to build similar high-transparency scientific products for other types of astronomical sources.\\
%such as e.g. Gamma Ray Bursts (GRBs) and astrophysical neutrinos.\\

%[By Jorge]
\noindent {\footnotesize\underline{Disclaimer.} The views expressed herein are those of the authors and do not necessarily reflect the views of the United Nations.}

\begin{acknowledgements}
     
\textbf{PG} acknowledges the support of the Technische Universit\"at M\"unchen - Institute for Advanced Studies, funded by the German Excellence Initiative (and the European Union Seventh Framework Programme under grant agreement no. 291763).

\textbf{CHB} acknowledges the support of ICRANet and the Brazilian government, funded by the CAPES Foundation, Ministry of Education of Brazil under the project BEX 15113-13-2.

\textbf{UBdA} acknowledges the support of a CNPq Productivity Research Grant no. 310827/2016-7 and a Serrapilheira Institute Grant number Serra - 1812-26906. He also acknowledges the receipt of a FAPERJ Young Scientist Fellowship.

\textbf{AVP} and \textbf{OC} are supported by the National Research Council of Argentina (CONICET) by the grant PIP 616, and by the Agencia Nacional de Promoci\'on Cient\'ifica y Tecnol\'ogica (ANPCYT) PICT 140492. \textbf{AVP} and \textbf{OC} are members of the Scientific Research career of the CONICET. This work is part of the PUE -IFLP (CONICET).
\end{acknowledgements}

%{\color{red}
%TO DO LIST \\

%\begin{itemize}
%\item Prepare the FITS, ASCII/csv and Web interactive table versions of the catalogue
%\item Include the catalogue into the SSDC searched %catalogue button in openuniverse.asi.it
%\item Introduce a new button in openuniverse.asi.it that %gives access to XRT images, similarly to "supercosmos" %and "radio surveys"
%\item Integrate the catalogue into VOU-Blazars/VOU-SED 
%\item Add the catalogue to the BSDC VO web page.
%\item Give access to XRT images (also in Hips format?) %associated to OUSXB sources form the web 
%\item Complete the list of highly monitored blazars
%\item other
%\end{itemize}
%}

%\begin{thebibliography}{}
\bibliographystyle{aa}
\bibliography{1ousxb}

% =========================
% APPENDIX
% =========================
\appendix
\section{The \swiftdeepsky\ pipeline}\label{pipeline}\label{appendix}
\swiftdeepsky\ is a software pipeline built on top of the Swift X-Ray Telescope Data Analysis Software (XRTDAS), designed and developed under the responsibility of the Space Science Data Center (SSDC) of the Italian Space Agency (ASI) and distributed within the NASA HEASoft\foothref{https://heasarc.gsfc.nasa.gov/docs/software/lheasoft/} data reduction package and related Swift-XRT calibration files (CALDB)\foothref{https://heasarc.gsfc.nasa.gov/docs/heasarc/caldb/}, that performs a complete X-ray image analysis of Swift XRT data.
A detailed description of the pipeline is given in Brandt et al. 2019 (in preparation); in this appendix we outline the main features.
 
\swiftdeepsky\ executes in sequence the following main tasks

\begin{enumerate}
\item Retrieve the list of all observation\_ID performed by Swift in the chosen period and select those executed in PC mode with exposure time larger than 200 seconds;
\item Download the data and calibration files that are necessary for the analysis from one of the Swift official archives;
\item Build exposure maps for every observation\_ID;
\item Add exposure maps and event files if more than a single observation\_ID has been performed in the chosen period (one week in this paper); 
\item The XIMAGE tool reads the event list and exposure map, builds X-ray images in the full 0.3-10.0 keV band;
\item The XIMAGE/DETECT command searches the X-ray image for pointlike sources;
\item XIMAGE estimates the background in four energy intervals:  the full band (0.3-10.0 keV), the soft (0.3-1.0 keV), medium (1.0-2.0 keV) and hard band (2.0-10.0 keV);
\item For every detected source in the full band, the XIMAGE/SOSTA tool estimates count-rates in all bands for all sources found by DETECT, and calculates three sigma upper limits in case a source is below the detection threshold;
\item NH command is run to estimate the amount of Galactic Hydrogen column in the direction of the source;
\item An estimate of a power law spectral slope is obtained by inverting count-rate ratio in the soft+medium and in the hard band;
\item Fluxes in all energy bands are calculated on the basis of count rates and NH assuming 
a power law spectrum with slope estimated from the hardness ratio or assuming an energy index of 0.8 for sources with less than $\sim$50 net counts;
\item Galactic absorption corrected \nufnu\ fluxes, suitable for SED plotting are calculated from the counts estimated in the soft, medium and hard band, and from the full band;
\item A second value of the power law spectral slope is estimated by means of a linear least square fit to the 0.5, 1.5 and 4.5 keV \nufnu\ fluxes;

\item If requested all the products are uploaded to a central database and made publicly available.
%{\color{magenta} comments by Sara T.: who/which data centre hosts the central database? which are the security measures that have been implemented so far to avoid (both unintentional and intentional) misconduct in uploading stuff to the central database?}
%{\color{red} P.G. at the moment the site is the BSDC-VO site hosted at ICRANet-Pescara, Italy. It needs to be improved in the coming months.}
\end{enumerate}

The software is publicly available either as open source code as well as  encapsulated in a Docker container ready to used in multiple platforms, MacOS, Linux and Windows10 Pro.

The source code and documentation to install and run the pipeline are available on Github:

\begin{itemize}
    \item \href{https://github.com/chbrandt/swift_deepsky}{\color{blue}{https://github.com/chbrandt/swift\_deepsky}}.
\end{itemize}

The Docker images are available from the Docker cloud:
\begin{itemize}
    \item \href{https://hub.docker.com/r/chbrandt/swift_deepsky}{\color{blue}{https://hub.docker.com/r/chbrandt/swift\_deepsky}}
\end{itemize}

As part of the processing, the exposure-maps of each observation are processed in real-time which demands the use of Swift (XRT) CALDB. CALDB is a rather big data package, and also evolves almost independently from the HEASoft tools. In the \swiftdeepsky\ such modularity is respected and what we do is to provide two containers that work together -- one for the pipeline and one for CALDB.

Once Docker is installed\foothref{https://www.docker.com/get-started} all that is needed to have the whole pipeline working is to \emph{pull} the \texttt{chbrandt/swift\_deepsky} and the \texttt{chbrandt/heasoft\_caldb} containers:
\begin{verbatim}
$ docker pull chbrandt/swift_deepsky
$ docker pull chbrandt/heasoft_caldb
\end{verbatim}

The pipeline is then run as a combination of two containers. 
To do that first instantiate the CALDB container,
\begin{verbatim}
$ docker run -dt --name caldb chbrandt/heasoft_caldb
\end{verbatim}
then run the pipeline:
\begin{verbatim}
$ docker run -it --rm --volumes-from caldb \
         chbrandt/swift_deepsky
\end{verbatim}
Since in this case no sky position was specified, the pipeline will simply output the command-line help message.

When a complete input is provided the results of the pipeline are written internally (to the container) to a directory called \texttt{/work}, this directory must be made available to any path in the host system. The following examples will use the users' home \texttt{tmp/sds} path to share the data.

For example, to run the pipeline around the position $(RA=317.58788, Dec=-86.2894)$:
\begin{verbatim}
$ docker run -it --rm --volumes-from caldb \
         -v $HOME/tmp/sds:/work \
         chbrandt/swift_deepsky \
              --ra 317.58788 --dec -86.2894
\end{verbatim}
another possibility is to give the name of a known astronomical object. In this case, the pipeline will consult Simbad\foothref{http://simbad.u-strasbg.fr/simbad/sim-fid} to get the corresponding coordinates.
Time windowing is also a possibility with \swiftdeepsky
\begin{verbatim}
$ docker run -it --rm --volumes-from caldb \
         -v $HOME/tmp/sds:/work \
         chbrandt/swift_deepsky \
              --object 3c279 \
              --start 2018-06-01 \
              --end 2018-12-31
\end{verbatim}
will use all Swift-XRT observations including the well known blazar $3c279$ taken from between 01/06/2018 and 31/12/2018.

Links to the pipeline software and documentation are available from the Open Universe portal (\href{http://openuniverse.asi.it}{\color{blue}{openuniverse.asi.it}}) under ``Open Software'' menu.

The Docker version of the \swiftdeepsky\ pipeline can be considered a "high-transparency" X-ray data analysis software tool because a) it can be downloaded and installed in a few clicks, b) it removes platform dependencies and the need to download the data and calibration files from the archive, and c) it lowers the barrier to X-ray data analysis enabling users, with or without experience in Swift-XRT data analysis, to run data reduction software that generates reliable science-ready standard data products usable by everyone.
%{\color{magenta} comment by Sara T: I th,ink we should specify that it removes the need to manually dowload the data, as the Docker container is doing it for the user - DONE} 

\end{document}